\documentclass[iop]{emulateapj}
\usepackage{apjfonts}
\usepackage{amsmath}
\usepackage{natbib}
\newcommand{\kt}{\ensuremath{k{\mathrm{T}}_{\mathrm{in}}}}
\newcommand{\LB}{\ensuremath{L_{\mathrm{B}}}}

\newcommand{\LX}{\ensuremath{L_{\mathrm{X}}}}

\newcommand{\Msol}{\ensuremath{\mathrm{M}_{\odot}}}

\newcommand{\NH}{\ensuremath{N_{\mathrm{H}}}}

\newcommand{\Zsol}{\ensuremath{\mathrm{Z}_{\odot}}}

\newcommand{\arcm}{\ensuremath{^\prime}}
\newcommand{\arcs}{\ensuremath{^{\prime\prime}}}

\newcommand{\pcmsq}{\ensuremath{\cm^{-2}}}

\newcommand{\erg}{\ensuremath{\mbox{erg}}}

\newcommand{\cm}{\ensuremath{\mbox{cm}}}

\newcommand{\nm}{\ensuremath{\mbox{\nm}}}

\newcommand{\ps}{\ensuremath{\s^{-1}}}
\newcommand{\pyr}{\ensuremath{\yr^{-1}}}
\newcommand{\s}{\ensuremath{\mbox{s}}}
\newcommand{\yr}{\ensuremath{\mbox{yr}}}

\newcommand{\ergps}{\ensuremath{\erg~\ps}}
\newcommand{\flux}{\ensuremath{\erg~\ps~\pcmsq}}

\newcommand{\Msolpy}{\ensuremath{\Msol~\pyr}}

\newcommand{\cf}{{\it c.f.\ }}

\newcommand{\etal}{{et al.\thinspace}}

\newcommand{\CHANDRA}{\emph{Chandra}}
\newcommand{\EINSTEIN}{\emph{Einstein}}

\newcommand{\XMM}{\emph{XMM-Newton}}

\newcommand{\chisq}{\ensuremath{\chi^2}}

\newcommand{\OVIII}{O~\textsc{viii}}



\begin{document}

\title{The Spectral and Temporal Properties of Transient Sources in Early-Type Galaxies  \\}

\author{N. J. Brassington}
\affil{School of Physics, Astronomy and Mathematics, University of Hertfordshire, College Lane, Hatfield AL10 9AB}
\affil{Harvard-Smithsonian Center for Astrophysics, 60 Garden Street, Cambridge, MA 02138}
\email{n.brassington@herts.ac.uk}
\author{G. Fabbiano}
\affil{Harvard-Smithsonian Center for Astrophysics, 60 Garden Street, Cambridge, MA 02138}
\author{A. Zezas}
\affil{Harvard-Smithsonian Center for Astrophysics, 60 Garden Street, Cambridge, MA 02138}
\affil{Physics Department, University of Crete, GR-710 03 Heraklion, Crete, Greece} 
\affil{IESL, Foundation for Research and Technology, 711 10 Heraklion, Crete, Greece}
\author{A. Kundu}
\affil{Eureka Scientific, Inc. 2452 Delmer Street, Suite 100, Oakland, CA 94602}
\author{D.-W. Kim}
\affil{Harvard-Smithsonian Center for Astrophysics, 60 Garden Street, Cambridge, MA 02138}
\author{T. Fragos}
\affil{Harvard-Smithsonian Center for Astrophysics, 60 Garden Street, Cambridge, MA 02138}
\author{A. R. King}
\affil{Theoretical Astrophysics Group, University of Leicester, Leicester LE1 7RH, UK}
\author{S. Pellegrini}
\affil{Dipartimento di Astronomia, Universita di Bologna, Via Ranzani 1, 40127 Bologna, Italy}
\author{G. Trinchieri}
\affil{INAF-Osservatorio Astronomico di Brera, Via Brera 28, 20121 Milan, Italy}
\author{S. Zepf}
\affil{Department of Physics and Astronomy, Michigan State University, East Lansing, MI 48824-2320}
\author{N. J. Wright}
\affil{Harvard-Smithsonian Center for Astrophysics, 60 Garden Street, Cambridge, MA 02138}

\shorttitle{Transient Sources in Early-Type Galaxies}
\shortauthors{Brassington \etal}
\bigskip

\begin{abstract}

We report the spectral and temporal variability properties of 18 candidate transient and potential transient (TC and PTC) sources detected in deep multi-epoch \CHANDRA\ observation of the nearby elliptical galaxies, NGC 3379, NGC 4278 and NGC 4697. Only one source can be identified with a background counterpart, leaving 17 TCs + PTCs in the galaxies. Of these, 14 are in the galaxy field, supporting the theoretical picture that the majority of field X-ray binaries (XRBs) will exhibit transient accretion for $>$75\% of their lifetime. Three sources are coincident with globular clusters (GCs), including two high-luminosity candidate black hole (BH) XRBs, with \LX=5.4$\times10^{38}$\ergps, and \LX=2.8$\times10^{39}$ \ergps, respectively.

The spectra, luminosities and temporal behavior of these 17 sources suggest that the transient population is heterogeneous, including neutron star (NS) and BH XRBs in both normal and high-rate accretion modes, and super soft sources containing white dwarf binaries. Our TC and PTC detections are noticeably fewer that the number expected from the populations synthesis (PS) models of \citet{Fragos_09}, tailored to our new \CHANDRA\ pointings of NGC 4278. We attribute this discrepancy to the PS assumption that the transient population is composed of NS XRBs, as well as differences between the statistical analysis and error estimates used in the model and our observations.

\end{abstract}


\section{Introduction}

\CHANDRA\ observations of elliptical and S0 galaxies have led to the widespread detection of low-mass X-ray binary (LMXB) populations \citep[see review][]{Fabbiano_06a}. The existence of these populations had been inferred from the first X-ray observations of early-type galaxies with the \EINSTEIN\ {\em Observatory} in the 1980s (Trinchieri \& Fabbiano 1985; see \citealt{Fabbiano_89} and references therein), but their first detection required the sub-arcsecond resolution of \CHANDRA\ \citep{Sarazin_00}. These populations of extra-galactic LMXBs provide a wider observational base for the understanding of LMXB formation and evolution, which has been debated since their discovery in the Milky Way (see e.g. \citealt{Giacconi_74}; \citealt{Grindlay_84}; review by \citealt{Verbunt_95}; \citealt{Piro_02}; \citealt{King_02}; \citealt{Bildsten_04}). A key observational result that can constrain the nature of these LMXBs is their time variability, as luminous LMXBs (\LX $\ge$ 10$^{37}$ \ergps) in relatively wide orbits are expected to exhibit transient behavior, according to the `standard' formation scenario of LMXBs in galactic fields \citep{Verbunt_95,Shahbaz_97,Piro_02,King_02,Wu_10}. For sources originating as ultra-compact binaries formed in Globular Clusters (GC) instead, transient behavior is only expected at lower luminosities \citep{Bildsten_04}. Until recently, only the most luminous extra-galactic LMXBs (\LX $\ge$ a few 10$^{38}$ \ergps\ in the 03$-$8.0 keV band) have been observed with \CHANDRA, with limited time coverage, making the detection of transients problematic. Recent deep monitoring observations of the nearby elliptical galaxies NGC 3379 (\citealt{Brassington_08}; from hereafter B08),  NGC 4278 (\citealt{Brassington_09}; hereafter B09) and NGC 4697 (\citealt{Sivakoff_08a}; \citealt{Sivakoff_08b}; hereafter S08) provide the means necessary to pursue this line of investigation.

Each of these galaxies have been observed with \CHANDRA\ as part of monitoring campaigns; NGC 3379 and NGC 4278 were observed as part of a legacy study of LMXB populations, with two new observations of NGC 4278 performed in 2010 also included (PI: Fabbiano). Multi-epoch observations of NGC 4697 were obtained from the archive. All three galaxies have been determined to be old $\sim$10 Gyr\footnote{\citet{Terlevich_02} estimated ages for the central galactic region for all of three galaxies, based on spectral line indices calibrated to the Lick system, via interpolation with the results of the \citet{Worthey_94} and \citet{Worthey_97} single stellar population evolutionary tracks. Further, for NGC 3379 and NGC 4278 SAURON data within the central region, again calibrated with the Lick indices and single stellar population synthesis models of \citet{Schiavon_07}, were used to determine age estimates \citep{Kuntschner_10}. For NGC 4697 \citet{Rogers_10} used spectral energy distributions of the stellar light, as well as absorption lines, which were then both compared to four different population models. This results in age ranges of: 9.3$-$14.7 Gyr, 10.7$-$14.1 Gyr and 8.1$-$11.0 Gyr for NGC 3379, NGC 4278 and NGC 4697 respectively.}, ensuring that the X-ray point source population is not contaminated by younger sources (e.g. high-mass X-ray binaries or supernova remnants).
In this paper distances of 10.6 Mpc (NGC 3379), 16.1 Mpc (NGC 4278), and 11.8 Mpc (NGC 4697) are adopted. These are based on the surface brightness fluctuation analysis by \citep{Tonry_01}. At these distances, 1\arcm\ corresponds to 3.1 kpc, 4.7 kpc, and 3.4 kpc, respectively.
A summary of the properties of the galaxies and their \CHANDRA\ observations is presented in Table \ref{tab:prop}, where column (1) gives the galaxy name, (2) distance, (3) the B-band luminosity, (4) the average stellar age, derived from single stellar population models, (5) globular cluster specific frequencies, (6) the \CHANDRA\ observation ID, (7) the date of each pointing, (8) the cleaned exposure time and column (9) the number of point sources detected in each observation within the overlapping area covered by all pointings. In this table, in addition to the single pointings, information from the coadded (All) observations are also provided.

The catalogs of point sources detected in NGC 3379 and NGC 4278 are presented in B08 and B09 respectively. The point sources of NGC 4697 presented in this paper were detected and analyzed using the techniques detailed in B08, all of these sources are included in the catalog presented in S08. A study of the X-ray luminosity functions (XLFs) of these galaxies in both the field and in GCs is presented in \citet{Kim_09}. \citet{Brassington_10} and \citet{Fabbiano_10} provide a detailed analysis of the bright persistent point source population of NGC 3379 and NGC 4278 respectively. PS modeling of the XLFs  and transient population were pursued by \citet{Fragos_08} and  \citet{Fragos_09}. The new point sources detected in the 2010 observations of NGC 4278 have been analyzed using the techniques presented in B09, and will be presented in a forthcoming paper. 

This paper is structured as follows: In Section \ref{sec:analysis} we describe the definition of transient candidate (TC) and potential transient candidate (PTC) sources, provide an overview of the properties of these sources from the B08 \& B09 catalogs, and our new analysis of NGC 4697, and present our results. In Section \ref{sec:results} we present our discussion of these sources along with comparisons to population synthesis modeling. Our conclusions are summarized in Section \ref{sec:conclusions}.

\section{Analysis and Results}
\label{sec:analysis}

\subsection{Transient Candidate Selection}
\label{sec:selection}

A transient candidate is usually defined to be a source that either appears or disappears, or is visible for only a limited amount of `contiguous' time during observations, where the flux ratio between the peak `on-state' emission and the non-detection upper limit, or `off-state' is greater than a certain value (usually between 5$-$10, e.g. \citealt{Williams_08}). However, using only this ratio as a discriminator can lead to overestimating the number of transients by including lower luminosity sources, which have poorly constrained \LX\ values. Following B08 and B09, we use instead the Bayesian model developed by \citet{Park_06}(see Section 2.4 of B08 for details), which estimates the uncertainties in the ratio, as well as lower-bounds\footnote{This lower-bound value accounts for the uncertainties in the source counts and upper limit values from the individual observations and thereby provides the lowest measure of `on' to `off' values for each source}. 
We classify sources with a lower bound ratio >10 as TCs and sources with ratios between 5 and 10 as PTCs.  We note here that a small number of Galactic sources have been observed to vary by ratios $>$10 but do not go into periods of quiescence, e.g. 4U 1705-440 \citep{Homan_09} and 4U 0513-40 \citep{Maccarone_10b}, where we particularly note that this latter source has been confirmed to be an ultra-compact binary \citep{Zurek_09}.

\begin{figure*}[t]
  \centering

    \includegraphics[height=0.6\linewidth]{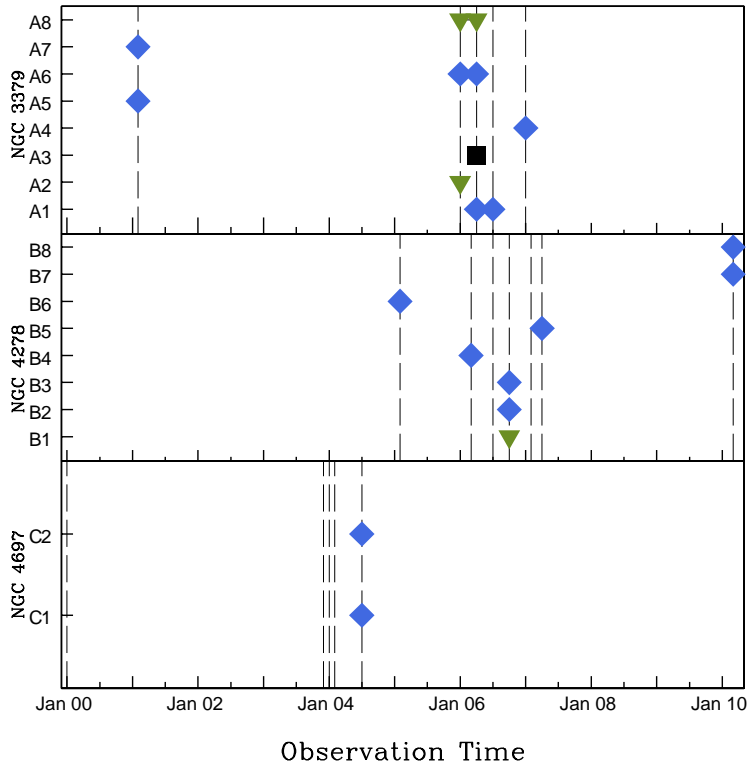}

    \caption{Long-term light curves for all the transient sources detected in the three galaxies, compared over the same time period. The x-axis indicates the observation date, with each pointing date indicated by a vertical dashed line in each panel, and the y-axis corresponds to the source numbers in Table \ref{tab:tcs}. Green triangles indicate confirmed GC sources, blue diamonds confirmed field sources and the black square indicates a background object. Source B7 is observed in 2 pointings, both of these observations were performed in March 2010 and hence only appear as one point in this figure. \label{fig:LC} }

\end{figure*}

\begin{figure*}[t]
  \centering
	\begin{minipage}{0.44\linewidth}
	\centering

    \includegraphics[width=\linewidth]{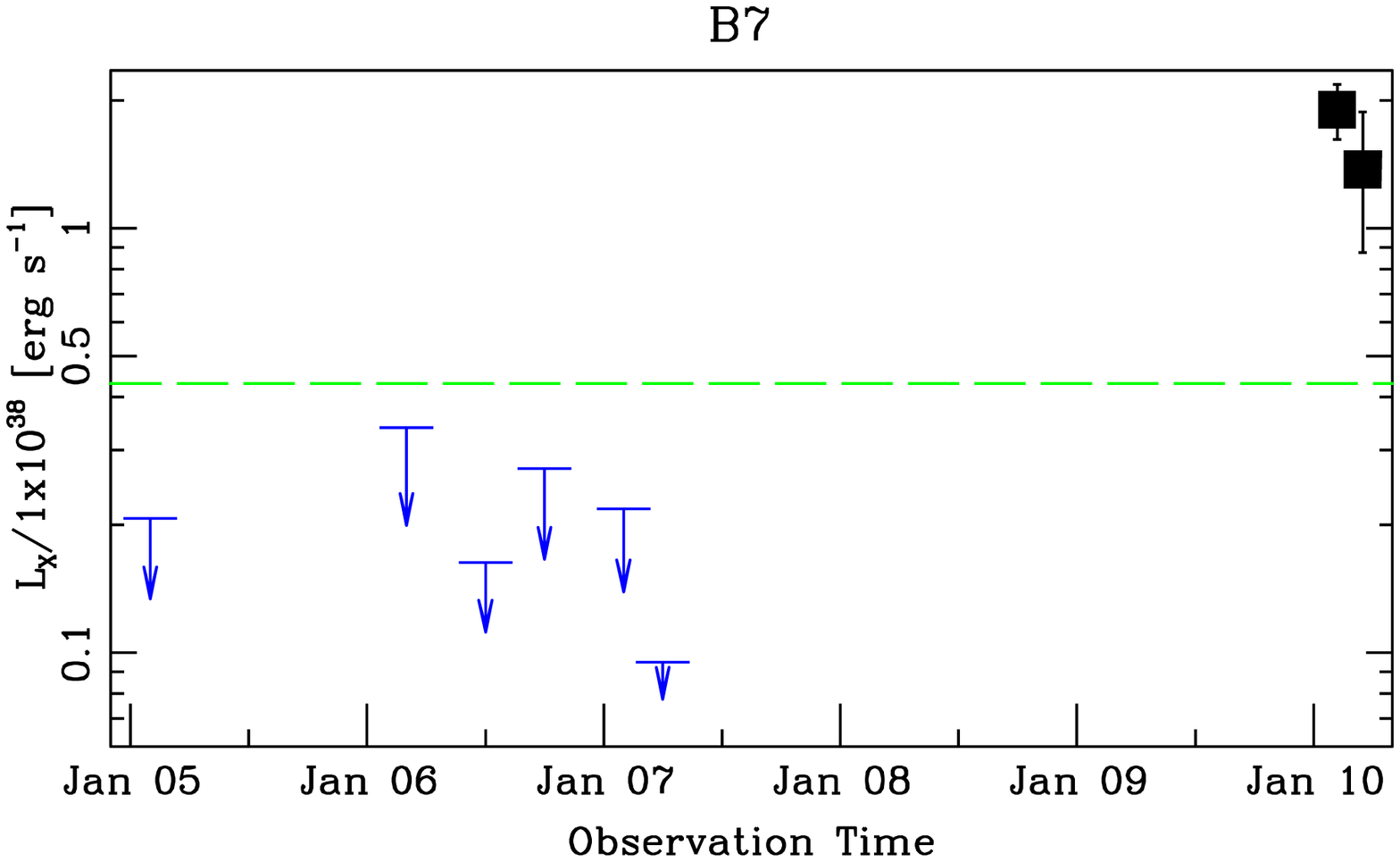}

	\end{minipage}
	\begin{minipage}{0.44\linewidth}
	\centering

    \includegraphics[width=\linewidth]{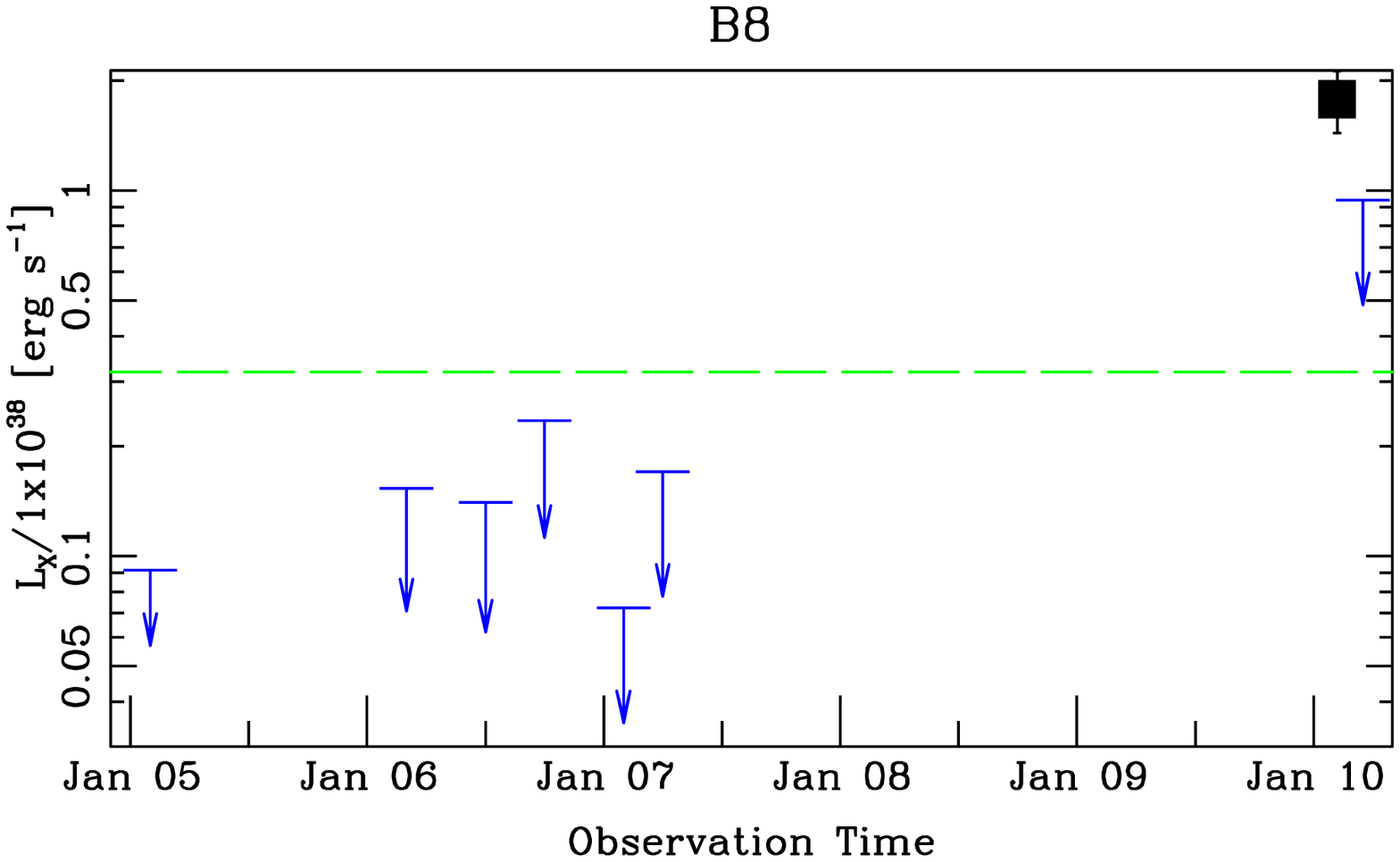}

	\end{minipage}

	\begin{minipage}{0.44\linewidth}
	\centering

    \includegraphics[width=\linewidth]{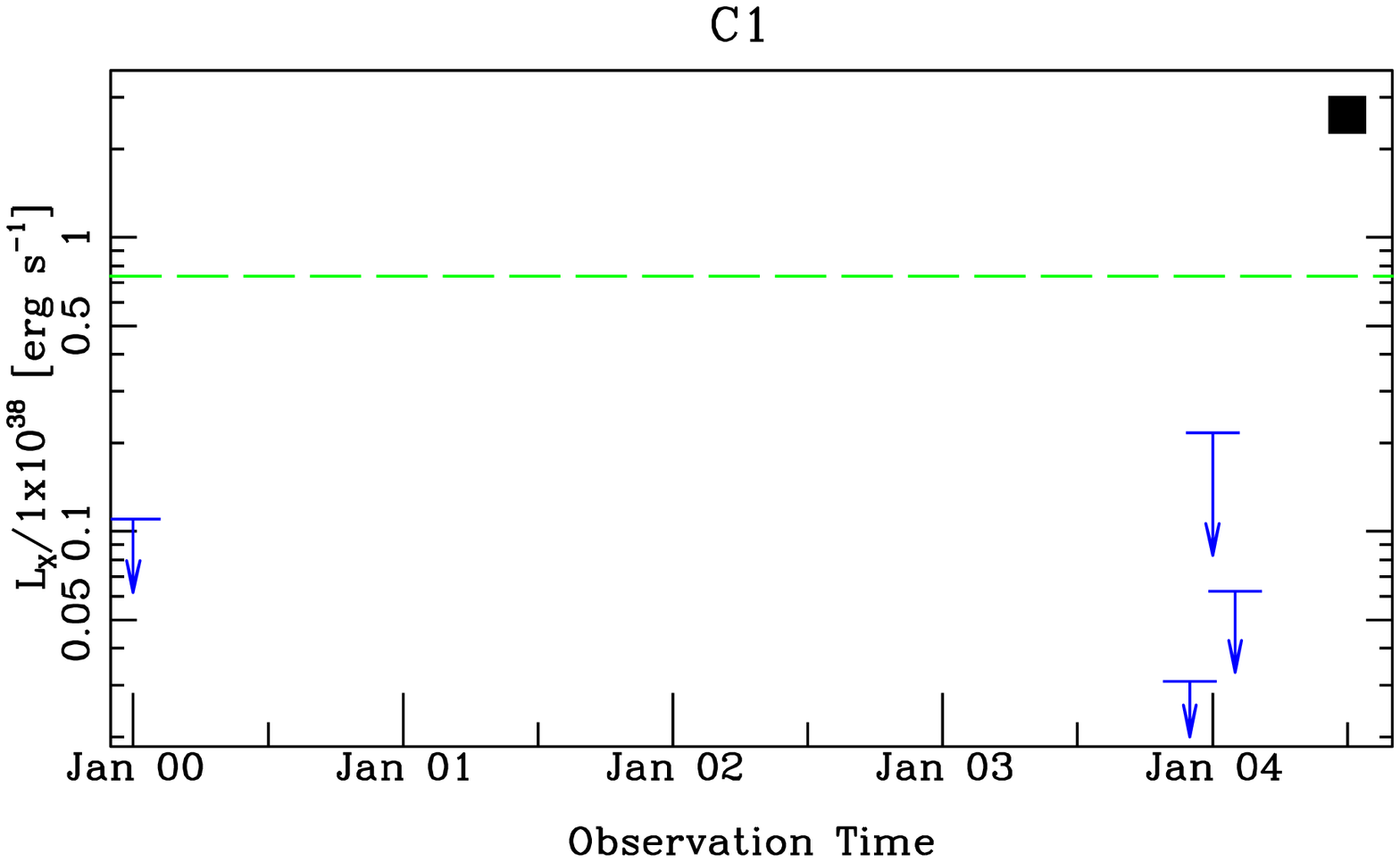}

	\end{minipage}
	\begin{minipage}{0.44\linewidth}
	\centering

    \includegraphics[width=\linewidth]{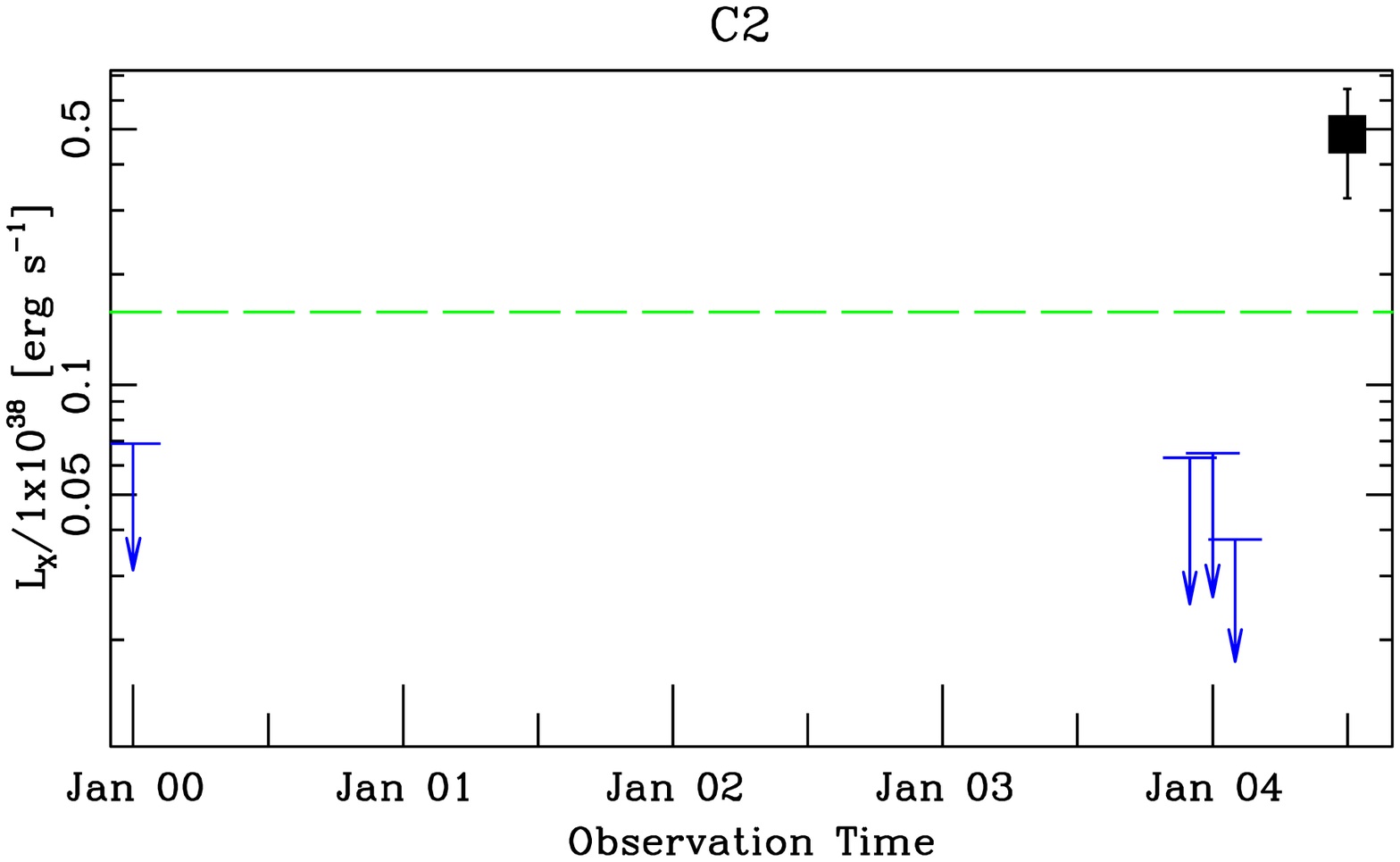}

	\end{minipage}

    \caption{Long-term light curves of the four additional sources that are not presented in B08 or B09. In cases where the source was not detected in an observation an upper limit of the X-ray luminosity has been calculated (details of this calculation are presented in Section 2.1 of B08). The horizontal line indicates \LX\ derived from the coadded observations. \label{fig:newLC} }

\end{figure*}

From the multi-epoch observations of the galaxies presented, the Bayesian analysis leads to the detection of 18 TCs and PTCs, listed in Table \ref{tab:tcs}. In this table column (1) indicates the source numbering used in this paper, and (2) indicates the original source number as presented in each catalog paper (B08, B09, S08 or `new' for sources detected in the NGC 4278 cycle 11 observations). Columns (3) and (4) provide RA and Dec of each source, (5) indicates the observation ID(s) the source was detected in. (6) provides net counts for that pointing (0.3$-$8.0 keV), (7) the mode ratio and (8) the lower-bound ratio from the Bayesian analysis, (9) marks the source as a TC or PTC, (10) the long-term light curve behavior (see Section \ref{subsec:var} for an explanation). In column (11) additional notes for each source are provided and (12) provides $g$ band magnitude (Vegamag) optical upperlimits for sources that have no detected optical counterpart. These upperlimit values are the 3$\sigma$ flux values centered on the RA and Dec values in columns (3) and (4). In B08 there was no optical coverage for source A8 but from recent {\em HST} observations it can now be confirmed that this source is coincident with a GC (with a separation $<$0.1\arcs), with  $g$ and $z$ band (Vegamag) magnitudes of 23.4 mag and 21.9 mag respectively. This corresponds to a color of 0.8 (ABMAGs), which indicates that the cluster is blue\footnote{These values and the upper limit values presented in column 12 of Table \ref{tab:tcs} have been derived using the methods presented in \citet{Kundu_07}.}. 
The long-term light curves are shown in Figure \ref{fig:LC} where the top, middle and bottom panels present the TCs and PTCs in NGC 3379, NGC 4278 and NGC 4697, respectively (In Figure \ref{fig:newLC} individual long-term light curves are presented for the four additional sources that are not presented in B08 or B09). 

In summary, we observe 5 TCs and 3 PTCs in NGC 3379, 3 TCs and 5 PTCs in NGC 4278, and 1 TC and 1 PTC in NGC 4697. Using the {\em HST} WFPC2 and ACS observations of these galaxies we find that three transients reside in GCs, and 14 in the galaxy stellar field. The source A3 in the NGC 3379 field, is found to be coincident with a background object; we only list this source and report its properties for completeness.

\begin{figure}
  \centering

    \includegraphics[height=1\linewidth,angle=-90]{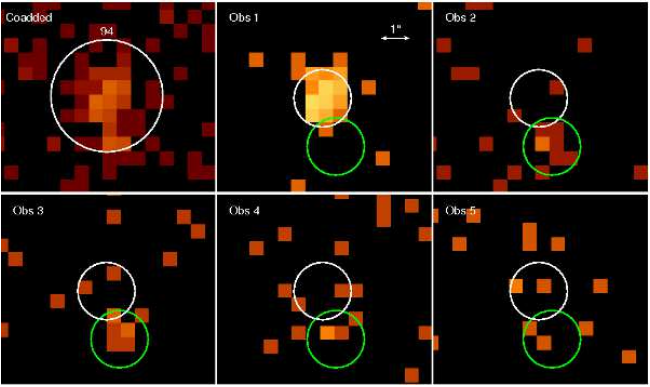}

    \caption{Full band (0.3$-$8.0 keV) \CHANDRA\ images of Source A5 from the coadded and individual observations. The white region in the top left panel indicates the 95\% encircled energy radius (at 1.5 keV) for A5 (labeled as S94 in B08), which was flagged as a possible double source in B08. In the subsequent panels the white region indicates the {\em wavdetect} position from observation 1 (obs ID 1587) and the green region indicates the {\em wavdetect} position from observation 2 (obs ID 7073). Both regions have a radius of 1\arcsec. The separation between these two positions is 1.86\arcsec\ (Astrometry offsets $\le$0.12\arcsec\ for NGC 3379; see Section 2.2 in B08). More details are provide in Section \ref{subsec:var}. \label{fig:src94} }

\end{figure}

\subsection{Flux Variability}
\label{subsec:var} 

The long-term light curves (Figure \ref{fig:LC}) show three TCs with `on states' of at least three months in NGC 3379. Of these sources, one has a maximum outburst time of 9 months (A1), while the other two (A6 \& A8) could have maximum outburst times up to 4 years and 5 months, respectively.  In NGC 4278, only source B7 was observed in two epochs (both in March 2010), with a minimum outburst time of $\sim$5 days and unconstrained maximum outburst time; the other TCs and PTCs were detected in single epochs, giving minimum outburst times of hours. Upper limits to the duration of the outbursts ranging from months to a few years can be placed on seven TC/PTCs, while the remaining seven were detected at the beginning or end of the monitoring observations and therefore the duration of the outburst is unconstrained. In Table \ref{tab:tcs}, column 10 lists our estimated outburst durations.

Source A5 is denoted as only being in outburst in one observation in Table \ref{tab:tcs}, although in B08 it was reported as being detected in three out of five observations. However, A5 was also flagged as a possible double source in B08, a conclusion supported by our further investigation (see Figure \ref{fig:src94}). The counts from observation 1 arise almost exclusively from within the top white circular region shown in this figure, whereas in the two subsequent observations the counts are centered in the lower green circle. The luminosity from observations 2 and 3 is more than a factor of 15 lower, suggesting that this secondary source is a variable lower flux object. 

Short-term variability was investigated for all sources with net counts$ > 20$ in a single observation, using both the Kolmogorov-Smirnov test (K-S test), and the Bayesian blocks method (BB) (\citealt{Scargle_98}; see Section 2.4 of B08 for more details). Of the 18 transient sources presented in this paper, only A5 and A8 exhibited detectable short-term variability during their outburst.

The intraobservation variability of these two sources was investigated in more detail by extracting the lightcurves from the `on' observations (and the results are discussed in sections \ref{subsec:srcFlare} and \ref{subsec:srcA8} respectively). 
This analysis was performed with the CXC CIAO software suite (v4.2)\footnote{http://asc.harvard.edu/ciao} and HEASOFT (v5.3.1), where barycenter corrected binned lightcurves were extracted with the CIAO tool {\em dmextract}. The binning for these two sources was selected to provide well constrained values, allowing variability to be identified, while remaining fine enough to ensure that any changes in flux were not masked by the selected binning scheme. Counts were extracted from circular source region files with radii as in B08. Background counts were extracted from a large elliptical source-free region, located within the overlapping area covered in all observations.

\subsection{Photometry}
\label{subsec:phot}

Hardness ratios (HR) and colors were derived for each source, as described in Section 2.3 of B08. They are listed in Table \ref{tab:phot} along with log \LX. These luminosity values were derived from the net count rate, assuming a power law shape of $\Gamma$=1.7 and Galactic \NH\ (see Section 2.1 in B08 for further details).

\subsection{Spectral Analysis}
\label{subsec:specanaly}

\begin{figure*}
  \centering
	\begin{minipage}{0.45\linewidth}
	\centering

    \includegraphics[width=\linewidth]{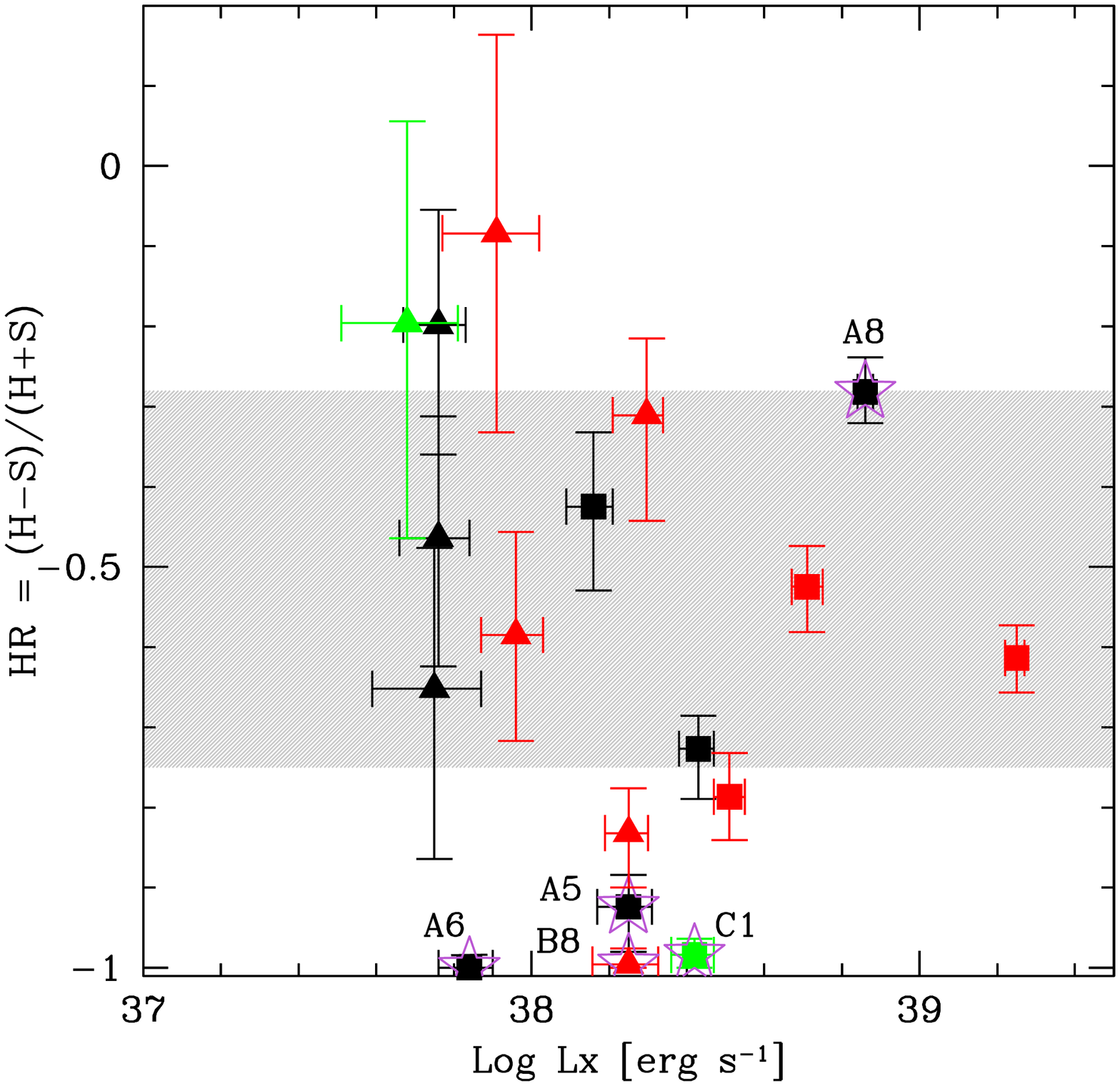}

	\end{minipage}
	\begin{minipage}{0.45\linewidth}
	\centering

    \includegraphics[width=\linewidth]{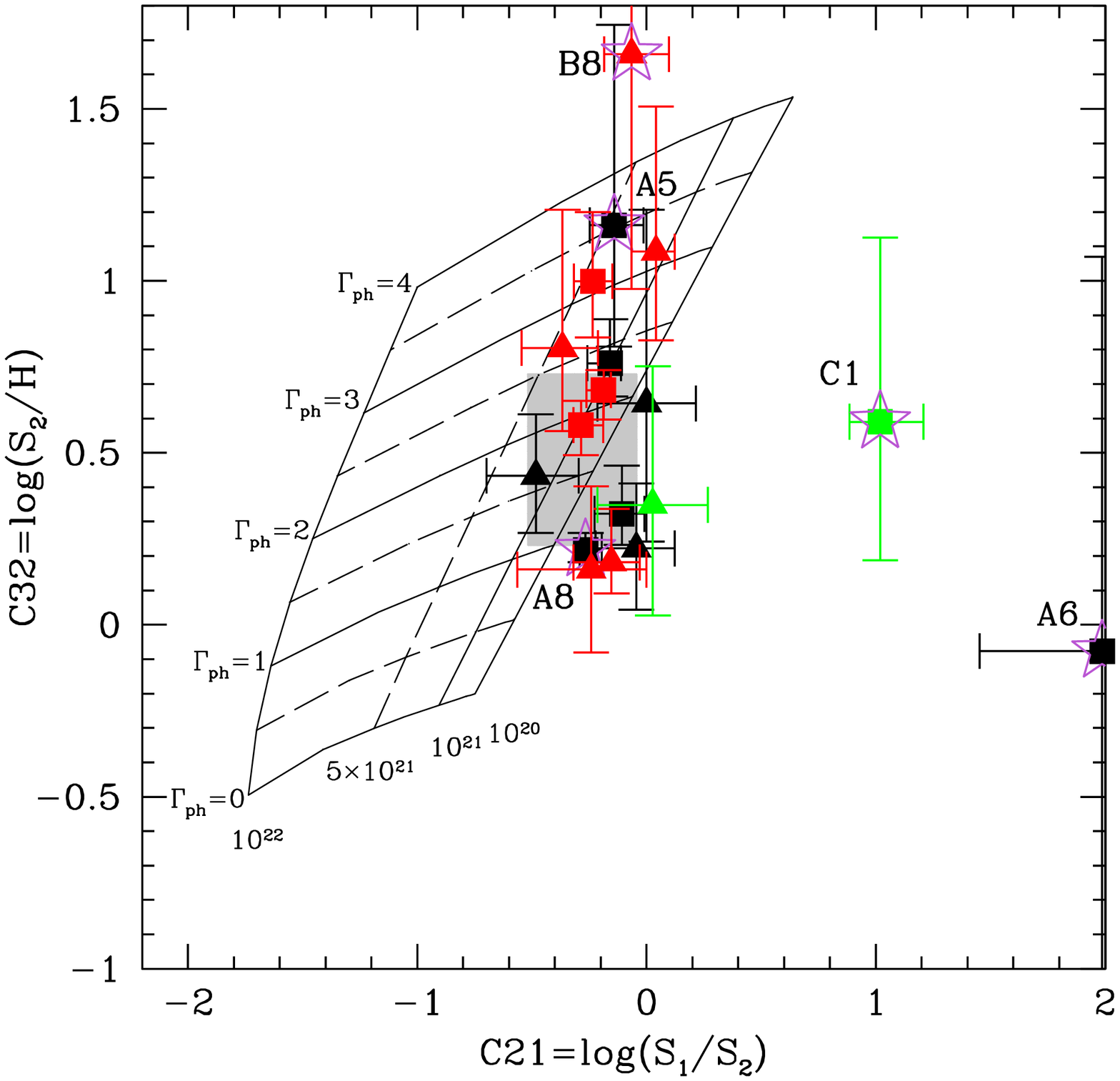}

	\end{minipage}

    \caption{Hardness ratio and color-color diagram of the Transient candidates in the three galaxies. Squares indicate TCs and triangles PTCs. Black points are sources from NGC 3379, red from NGC 4278 and green NGC 4697. The shaded regions indicate the range in values of 90\% of the persistent sources in the three galaxies (i.e. net counts $>$50 counts). Also shown in the right hand panel is a grid to indicates the predicted locations of the sources at redshift $z$=0 with various photon indices (0 $\le \Gamma_{\mathrm{ph}}\le 4$, {\em top to bottom}) and absorption column densities (10$^{20} \le \NH\ \le 10^{22}$ \pcmsq, {\em right to left}). The five TC/PTCs that have been identified to have unusual spectral properties (see section \ref{sec:results} for more details) are indicated by purple stars and are labeled in both panels. \label{fig:pop} }

\end{figure*}

Spectral extraction was performed for the `on' observations for the TC/PTC sources. using the CIAO tool {\em psextract}. We used the same circular extraction regions as in B08 or B09;  for the TC/PTC discovered in our new analysis, the radii are: 2.5\arcsec\ (B7), 3.0\arcsec (B8), 2.2\arcsec\ (C1) and 3.0\arcsec (C2); background counts were extracted from surrounding annuli, with outer radii 2-3 times larger, depending on the presence of nearby sources. In cases where the source region was found to overlap with a nearby source, the extractions region radius was reduced (to a minimum of 1.5\arcsec) and the area of the overlapping source was excluded.

The source spectra were fitted in XSPEC (v12.5.1). We used only data between 0.3$-$8.0 keV, to avoid the calibration uncertainties at the low energies, and the high cosmic ray background above 8.0 keV. All the spectra were fitted to two models that describe the properties of X-ray binary spectra well (see e.g. review by \citealt{Remillard_06}): the multicolor disc blackbody (DISKBB in XSPEC; hereon in referred to as DBB), and the power law (PO) models. For each model all parameters were allowed to vary freely, with the exception of the absorption column \NH, described by the X-ray absorption model {\em tbabs}, which in cases where the best-fit value was below that of the Galactic absorption, was frozen to the Galactic value\footnote{Galactic \NH\ was calculated with the tool COLDEN: \\ http://cxc.harvard.edu/toolkit/colden.jsp. This tool provides values of (2.79, 1.79 \& 2.14)$\times 10^{20}$\pcmsq\ for NGC 3379, NGC 4278 and NGC 4697 respectively.}.
In instances where there were sufficient counts to bin the data to at least 20 counts per bin (allowing Gaussian error approximation to be used), the minimum \chisq\ method was used to fit the data. The Cash statistic \citep{Cash_79} was otherwise used, however this statistical method has the disadvantage that it does not provide a goodness of fit measure like \chisq. 

Table \ref{tab:Bestfit} summarizes the best-fit models, with the three galaxies listed separately. Column (1) contains the source number, (2) the observations used in each fit, (3) the net source counts from the spectral extraction, (4) indicates which spectral model was used, (5) the fit statistic (\chisq\ and number of degrees of freedom $\nu$, or the C statistics - indicated by $C$), (6) the null hypothesis probability (or goodness when using Cash). Columns (7), (8) \& (9) present the best-fit values of the fit parameters with 1$\sigma$ errors for each interesting parameter (F denotes that the value was frozen): \NH, $\Gamma$ for the power law model and \kt, the temperature of the innermost stable orbit of an accretion disc, for the multicolor disc model. Column (10) indicates the intrinsic value of \LX\ and (11) the luminosity range from the lowest (1$\sigma$) lower bound to the highest (1$\sigma$) upper bound of each observation included in the joint fit\footnote{Throughout this paper luminosities quoted are the 0.3$-$8.0 keV \CHANDRA\ bandpass. $L_{\rm{bol}}$ corrections for the HS sources will be between 5$-$15 \citep{Portegies_04} and corrections for SSS/QSS will be between 1.5$-$3 (e.g. \citealt{McGowan_05}). Bolometric corrections for sources in a TD state are small, with $\sim$90\% of the emission occurring between 0.3$-$8.0 keV \citep{Portegies_04}.}.

\begin{figure*}
  \centering
	\begin{minipage}{0.4\linewidth}
	\centering

    \includegraphics[height=\linewidth,angle=-90]{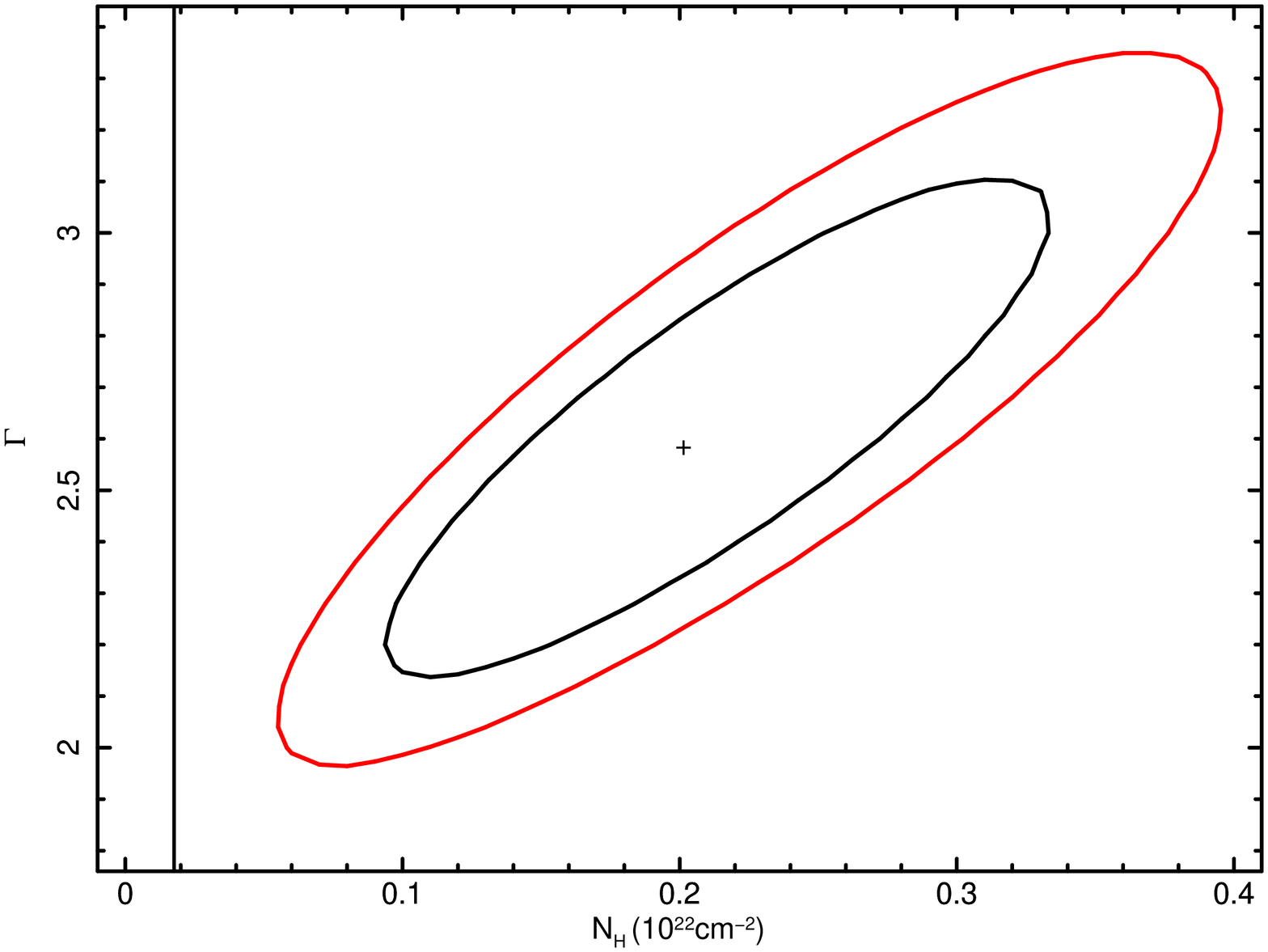}

	\end{minipage}
	\begin{minipage}{0.4\linewidth}
	\centering

    \includegraphics[height=\linewidth,angle=-90]{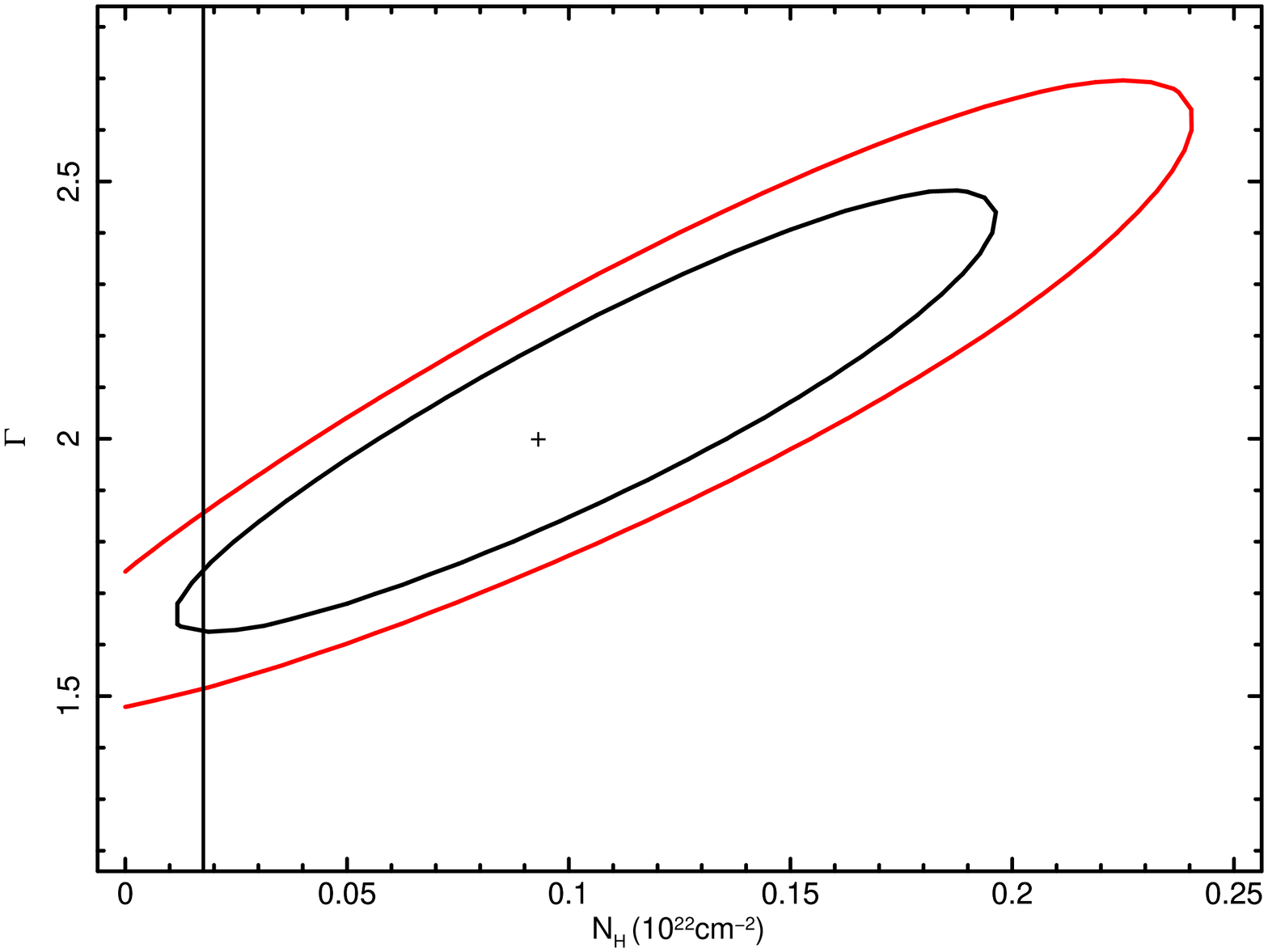}

	\end{minipage}

    \caption{1$\sigma$ and 2$\sigma$ contour plots for single-component PO models from sources A1 (left) and B6 (right), presented in Section \ref{subsec:specanaly}. In both figures the the cross hair within the contours indicates the best-fit values and Galactic \NH\ is denoted by the vertical line. \label{fig:conpo} }

\end{figure*}

Figure \ref{fig:pop} shows that the majority of TC/PTCs have both HR and colors  consistent with those of the persistent LMXBs in the galaxies. 
Five of these sources have insufficient counts to allow even simple spectral modeling (A2, A7, B5, C2 \& the background source A3, which will not be discussed further; see Table \ref{tab:Bestfit}) and therefore single-component models with canonical values have been applied ($\Gamma=1.7$ or \kt=1.0 keV \citealt{Remillard_06}). Six more sources (A1, A4, B1, B4, B6 \& B7) have spectra well described by single-component models. Four of these (A4, B1, B4 \& B7), are well fitted with a PO model with $\Gamma \sim$1.7 and Galactic \NH. A1, instead requires a large value of \NH\ in the PO fit. Based on the simulations of \citet{Brassington_10}, this result may favor disk emission. The single-component PO model of source B6 also shows an elevated value of \NH, however, the two-dimensional errors on this value indicate that the best-fit \NH\ is consistent with Galactic absorption (Figure \ref{fig:conpo}).

Sources B2 \& B3, have colors and HR values softer than the persistent LMXB population (Figure \ref{fig:pop}).  Their spectra are well fitted by both a single PO and single DBB model. However, the PO model in both cases requires an elevated value of \NH, indicative of the source containing a significant disc component \citep{Brassington_10}. In the case of B2, the best-fit value of \NH\ from the DBB model tends to zero when left free to vary, indicating that there is likely a non-thermal component as well as disc emission in this source \citep{Brassington_10}. When freezing the absorption column to the Galactic value, the best-fit disc temperature is 0.41 keV, with an X-ray luminosity of 1.0$\times10^{38}$\ergps. The spectrum of source B3 is well described by a DBB model with an absorption value more than twice that of the Galactic absorption value (albeit poorly constrained), indicating that more than 75\% of the emission is arising from the disc component, with a temperature of 0.55 keV.

Four further sources (A5, A6, B8 \& C1; see Figure \ref{fig:pop} to compare with the persistent sources), all have little emission above 2 keV, and are therefore super soft, or quasi-soft, sources (SSS and QSS respectively). In all cases, the PO fits yielded extremely steep power-law indices (>6), reflecting the very soft nature of the emission. Their spectra were also fitted with black-body models (BB), as reported in Table \ref{tab:Bestfit}.

When simple single-component models did not provide an adequate description of the source spectrum, we used more complex models, as discussed below for sources C1 (Section \ref{subsec:c1sss}) and A8 (Section \ref{subsec:srcA8}).

\section{Discussion}
\label{sec:results}

\subsection{The Field Transient Population}

One of the clear results from this work is that TC/PTC sources are predominantly found in the field with only three sources out of the 17 TC/PTCs (excluding the background source) coincident with a GC\footnote{In comparison observations of the total X-ray binary population suggest that between 30\%$-$70\% of LMXBs are coincident with a GC (e.g. \citealt{Angelini_01}, B08, B09). We also note that the optical coverage of all three galaxies provided a representative view of the GC population of each of the systems (see B08, B09 and \citealt{Kim_09} for further details).}. These numbers support the theory which suggests that the LMXB field population largely arises from relatively detached systems evolved from native field binaries and as such will exhibit transient accretion for $>$75\% of their lifetime \citep{Piro_02}. This is as opposed to transient GC-LMXBs, which have a more difficult formation channel (as discussed in Section \ref{subsec:GC}).

From the detailed spectral analysis presented in this paper we have been able to determine that out of the 14 confirmed field sources, 10 exhibit colors and spectra consistent with typical emission from LMXBs, with power-law ($\Gamma$ 1.3$-$2.0) or disk  (\kt\ 0.4$-$1.0 keV) emission and on-state intrinsic luminosities \LX\ ranging from 4$\times10^{37}$ \ergps\ to 2.32$\times10^{39}$ \ergps\ (see Table \ref{tab:Bestfit}). The spectra of these sources are consistent with those of NS or BH binaries in either a hard state (HS) or thermally dominant (TD) state \citep{Remillard_06}.

Only one of these sources has a luminosity significantly above the Eddington limit for a 1.4\Msol\ neutron star (B6) and is a strong BH candidate. A further three sources that have been determined to be in a TD state (A1, B2, B3) exhibit inner-disc temperatures that are softer than the spectra typically observed in Galactic NS-LMXBs (\kt\ 1$-$2 keV), indicating that they could be BH candidates. A further three field sources (A4, B4 \& B7) have been determined to be in a HS, but have X-ray luminosities $\ge$10\% of the Eddington limit. This high accretion is not observed in NS-LMXBs in a HS and suggests that the primary of these binaries may be BHs. However, due to the quality of our data, while a HS is the preferred interpretation, we cannot rule out any of these sources emitting in a TD state, and therefore cannot exclude any of these sources from being NS-LMXBs. The remaining three sources have low-count data and have \LX\ values derived from canonical models, therefore these sources could be NS or BH-LMXBs. A further soft source (B8) has been classified as a QSS but is also consistent with a NS-LMXB (or WD-LMXB; see section \ref{subsec:QSS}).
This results in a ratio of BH to NS-LMXBs in transient field sources ranging from at least 10\% up to 90\% of sources. Such an unconstrained ratio is consistent with a study of the transient population of M31 \citet{Williams_06} which suggest that the majority of the sources that they identify to be LMXBs contain BHs.

\subsubsection{Comparison with PS models of field LMXBs}

Population synthesis modeling of the LMXB populations of NGC 3379 and NGC 4278 were presented in \citet{Fragos_08} and a study comparing the theoretical and observed field transient populations of these galaxies was provided in \citet{Fragos_09}. From this work it was determined that a constant duty cycle (DC) for all transient systems did not match the data and instead a variable DC for each system provided results that were consistent with the observations. Furthermore \citet{Fragos_09} suggested that the observed number of TCs and PTCs is proportional to the total number of field LMXBs (or the stellar mass of the galaxy) and not the total observed LMXB population (which will be enhanced by GC-LMXB formation; \citealt{Kim_06}). 

These models were compared to the transient population of NGC 4278 prior to the 2010 observations and as such enabled us to estimate the number of new TC and PTC sources that would be observed with repeat pointings. Based on the models' prediction the additional 2010 observations of NGC 4278 would reveal an additional $\sim$6 TCs and $\sim$6 PTCs. However no TCs and only two new PTCs have been determined from these recent observations. 
The discrepancy between the model predictions and the observed number of new sources can be attributed to three main factors. \citet{Fragos_09} assumed in their modeling calibration that all TCs and PTCs determined from B08 \& B09 were NS-LMXBs. This assumption was based on the fact that NS-LMXBs were the dominant population in their population synthesis models, and the lack of detailed characterization of the observed transient population. From the spectral modeling that has been presented in this paper, it can now be determined that only $\le$2 TCs and $\le$1 PTC, and $\le$1 TC and $\le$3 PTCs  in NGC 3379 and NGC 4278 respectively are classified as NS-LMXBs (compared to 5 TC and 3 PTC, and 3 TC and 3 PTC previously). Combining this fact with the low number statistics used in the model-observation comparison by \citet{Fragos_09} can possibly explain the over-prediction of their models.

Furthermore, \citet{Fragos_09} used the `on-to-off' flux ratio of a source in order to characterize it as a TC or PTC, instead of the more conservative Bayesian analysis employed here. Finally, we should also consider the uncertainties of the bolometric corrections used by \citet{Fragos_08,Fragos_09} in order to convert mass-transfer rates to X-ray luminosities, which would significantly increase the error bars in the predicted number of transients. These last two factor can explain the discrepancy between  the model predictions and observations based solely on statistical arguments.

\subsection{GC-LMXBs}
\label{subsec:GC}

Three of the TC/PTCs presented in this work are in GCs. It has been suggested that there is a trend for GCs to have a higher fraction of ultra-compact binaries, compared to the field LMXB population \citep{Deutsch_00,Heinke_10}. However for sources brighter than $\sim \times10^{37}$\ergps\ such objects are expected to be persistent \citep{Lasota_08,Bildsten_04} and are therefore not the transient sources presented here. Instead these sources could be NS with main sequence or red giant donors, where population synthesis models of the formation of LMXBs in GCs predict that these systems, in addition to ultra-compact binaries, also contribute to the GC-LMXB population \citep{Ivanova_08}. Such sources were shown to exhibit transient emission from 60\% to all of their mass-transfer lifetimes. We therefore suggest that A2, a source with low counts which has therefore been described with a canonical PO spectrum, is likely to be a NS-main sequence or NS-red giant system with \LX$\sim8\times10^{37}$\ergps\ (although NS-red giants are less favored as \citet{Ivanova_08} predicted that their numbers are low when considering their formation rates and short lifetimes).

The second GC-LMXB, B1, has a spectrum that is well described by both a PO and DBB model, however the best-fit \LX\ from the PO model indicates that the source is emitting greatly above the 2$-$4 percent Eddington luminosity that is typically seen from a LMXBs in a HS \citep{Maccarone_03} and the DBB model is therefore preferred\footnote{It is also noted that while the absorption column of the PO model indicates a value consistent with the Galactic value, and therefore indicative of a HS (\cf\ \citealt{Brassington_10}), this value is poorly constrained and consequently does not rule out the presence of a thermal component.}. The best-fit values from the DBB model determines an inner disc temperature of \kt$\sim$=1.3 keV and \LX=5.4$\times10^{38}$\ergps.
Therefore, because of this high luminosity, which is above the Eddington limit for a NS binary (although could be close to the limit of a heavy (2$-$3 \Msol; \citealt{Kalogera_96}) NS, or a 1.4\Msol\ NS with a He or C/O donor), it is possible that B1 is a BH binary instead of a NS-main sequence system. 

The third GC-LMXB in this sample is A8 (presented in section \ref{subsec:srcA8}), which could be a BH-ULX undergoing a period of super-Eddington accretion.

The detection of a BH binary in a GC is thought to be rare, since in tidally captured persistent sources BHs are likely to be expelled from the GC \citep{Spitzer_69}; they could be transients with very low duty cycles from an exchange interaction \citep{Kalogera_04}. More recent work has indicated that a significant number of BH can be retained in GCs \citep{Mackey_07,Moody_09}. In particular, \citet{Maccarone_11} point out that once a number of BHs are ejected from their host GC the ratio of the masses between the heavy (BHs) and light (non-BH) components falls below the critical value and the Spitzer instability criterion is no longer met, therefore not all BHs will be expunged.

Observationally, the existence of accreting BHs in GCs has been supported by the detection of five luminous (above the NS Eddington limit), variable sources \citep{Maccarone_07, Brassington_10, Shih_10, Irwin_10, Maccarone_11}. Using the TC/PTC criteria defined in this paper, four of these binary systems are persistent and could be ultra-compact binaries with either a BH \citep{Gnedin_09} or NS with super-Eddington accretion and rather mild beaming \citep{King_11}\footnote{Although \citet{Peacock_12} noted that the model presented in \citet{King_11} is inconsistent with the observed emission of bright X-ray sources in GCs, where the observed X-ray emission is harder. They also suggested that many of the sources considered in \citet{King_11} will be dominated by carbon and oxygen edges instead of Thomson scattering, and therefore, as Thomson scattering is assumed in \citet{King_11}, this model cannot be applied.}.
The BH-GC candidates A8 and B1 exhibit large flux variability, with \LX$>1\times10^{38}$\ergps, therefore ruling out genuine disc instability transients. They are also unlikely to be the mildly beamed super-Eddington NS-LMXBs suggested by \citet{King_11}.

Instead, A8 and B1 could represent BHs in a GC that have been formed through an exchange interaction, resulting in transient behavior \citep{Kalogera_04}. In \citet{Barnard_11} this formation channel has been advocated to explain a recurring transient in M31, which has been observed with an outburst luminosity$\sim L_{\rm{Edd}}$ and therefore could be a BH-GC candidate (although a NS binary cannot be ruled out). In their work they also suggest that the source could have formed through a complex interaction of a triple system resulting in a BH-WD binary, as first theorised by \citet{Ivanova_10}. Further, additional observations of the first strong BH-GC candidate (first presented in \citealt{Maccarone_07}), have lead to the suggestion that this source is also a triple system, with the inner binary comprising a BH and WD \citep{Maccarone_10}. 

The properties of the host clusters for the five BH-GC candidates are presented in \citet{Maccarone_11}, where they demonstrate that this small population suggests that BH binaries are favored in massive `red' GCs. The properties of the GCs hosting A8 and B1 are also massive with $g$=23.4 and $V$=21.6 respectively. However, both of these sources have color values indicating that these GCs are blue, (although B6 has a ($V-I$) of 1.00$\pm$0.02 which is only marginally classified as a `blue' low metallicity cluster: \citealt{Fabbiano_10}; blue clusters have $V-I\le1.05$). This further indicates that the mass of the GC is important in the formation of GC-LMXBs in these environments. 
However, we do not comment on the influence of GC colors in the formation of BH-GCs, as the metallicity correlation presented in \citet{Maccarone_11} was only suggestive.

\subsection{A8: ULX in Outflow in a GC}
\label{subsec:srcA8}

\begin{figure}
  \centering

    \includegraphics[height=0.9\linewidth,angle=-90]{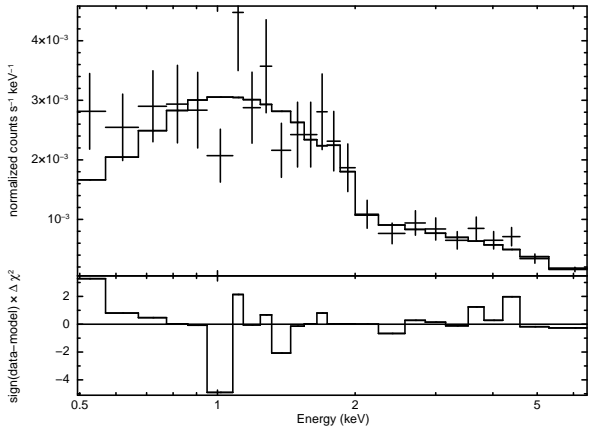}

    \caption{The whole spectrum of source A8 from observation 2 (obs ID 7073) indicating the PO model fit and the \chisq\ residuals, which indicate the significant deficiencies between 1$-$2 keV. The details of this model are discussed in Section \ref{subsec:srcA8}. \label{fig:posrc128} }

\end{figure}

The GC source A8 in NGC 3379 is very luminous (Table \ref{tab:Bestfit}) and variable (Section \ref{subsec:var}). The photometric parameters are consistent with those of the persistent LMXB population (Figure \ref{fig:pop}), and a single-component PO model provides a statistically acceptable description of the spectrum, with $\Gamma=$1.1 (Table \ref{tab:Bestfit}). However, there is a statistically significant deficiency in the spectrum at $\sim$1.0 keV as well as excess emission $<$0.7 keV (Figure \ref{fig:posrc128}). These features cannot be modeled with either a composite DBB plus PO model, or additional neutral absorption components, or a thermal plasma model. A single-component power law model, modified by an ionized intrinsic absorber ({\em absori}), was fitted to the spectrum. The results of this fit are presented in Table \ref{tab:spec128}, where the best-fit values from an XSPEC model with {\em tbabs$\times$tbabs$\times$absori$\times$po} (column 2) are shown with one of the neutral absorption components being frozen to Galactic \NH.

\begin{figure}
  \centering

    \includegraphics[width=\linewidth]{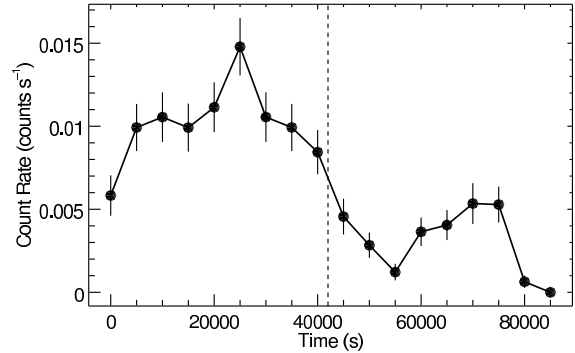}

    \caption{Short-term light curve of Source A8 from observation 2 (obs ID 7073), with binning of 5000s. The dashed vertical line indicates the end of the high flux period, which has a count-rate of $>$0.01 cnt s$^{-1}$. The lower emission state has a count-rate $<$0.005 cnt s$^{-1}$. We used a finer 500s binning to determine the onset of this decrease in flux, which occurs $\sim$42 ks into the observation. Separate spectra were extracted from before and after this time. This behavior is discussed in Section \ref{subsec:srcA8}. \label{fig:lc128} }

\end{figure}

\begin{figure*}
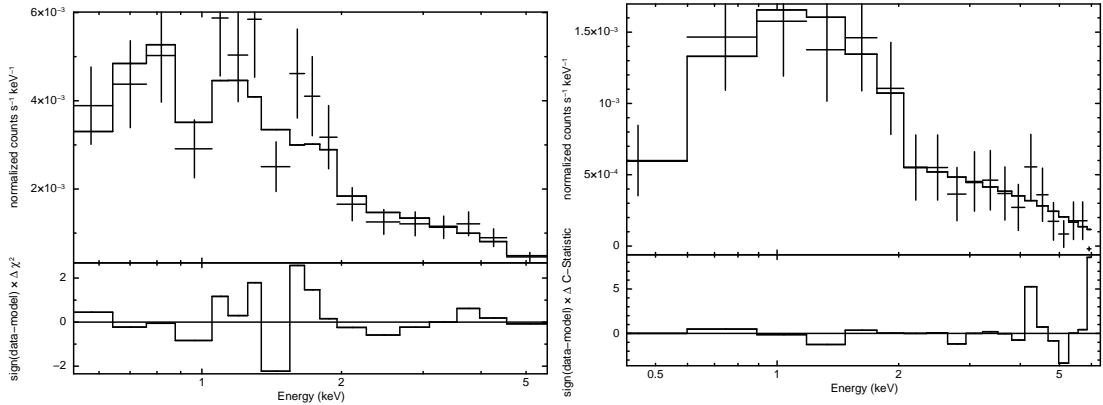

  \centering
	\begin{minipage}{0.4\linewidth}
	\centering

    \includegraphics[height=\linewidth,angle=-90]{f8a.ps}

	\end{minipage}
	\begin{minipage}{0.4\linewidth}
	\centering

    \includegraphics[height=\linewidth,angle=-90]{f8b.ps}

	\end{minipage}

    \caption{Left: Best-fit spectrum and \chisq\ residuals from the `high' flux emission of source A8 where an ionized absorption component has been included in the model. Right: Spectrum and Cstat residuals of the `low' emission period from A8 fitted to a single-component PO model. This is discussed in detail in Section \ref{subsec:srcA8}. \label{fig:src128} }

\end{figure*}

A8 is one of the two TC/PTC sources exhibiting short-term variability, where the count-rate falls from $>$0.01 cnt s$^{-1}$ to $<$0.005 cnt s$^{-1}$ $\sim$42 ks into the observation (see Figure \ref{fig:lc128}). The spectrum of the high flux state requires an ionized absorption component. The best-fit values of the absori power law model are presented in column 3 of Table \ref{tab:spec128}, and the spectrum is presented in the left hand panel of Figure \ref{fig:src128}. These values are similar to those determined for the spectrum from the full observation, except that the best-fit $\Gamma$ is 1.89 and \NH\ is larger (although both parameters are consistent with the values derived from the whole observation when considering the uncertainties). The derived X-ray luminosity for the source in this higher flux state is 2.8$\times10^{39}$ \ergps, in the {\em Ultraluminous X-ray Source} (ULX) range.

The spectrum of the lower state is instead adequately described by a single-component PO model (right hand panel Figure \ref{fig:src128}), with \NH\ consistent with Galactic (column 4 of Table \ref{tab:spec128}). However the best-fit photon index is very flat, similar to the value derived when fitting the whole spectrum to a single-component model. This could indicate that ionized absorption may still be present, although there are too few counts to be able to investigate it. The hardness ratios, binned to 5000s over this period, are constant. From this model a luminosity of 8.9$\times10^{38}$ \ergps\ is derived.

This source was also detected in the 3rd observation of NGC 3379 (OBS ID 7074), although it only has $\sim$9 net counts, too few to perform any meaningful spectral analysis. B08 estimated an X-ray luminosity of \LX=2.5$\times10^{37}$ \ergps. This luminosity then fell to an upper limit of 2$\times10^{37}$ \ergps\ in the subsequent pointing, taken three months later.

The parameters determined from the model during the period of high flux emission are similar to those reported for the NGC 1365 flaring ULX by \citet{Soria_07}. In A8, the properties of the spectrum (see Table \ref{tab:spec128}) suggest that the source is in some sort of outflowing phase, similar to the decline period of the ULX in NGC 1365. During this phase the spectrum is fitted with an ionized absorption column of $\sim5^{+3}_{-2}\times10^{22}$\pcmsq\ and an ionization parameter of $\xi\approx$150$^{+79}_{-52}$ \erg\ cm s$^{-1}$, largely consistent with the parameters determined in \citet{Soria_07}, of $\sim1^{+2}_{-1}\times10^{22}$\pcmsq\ and 111$^{+184}_{-83}$ respectively, where they also determine $\Gamma$ $\sim$1.9 and neutral absorption column $\sim2\times10^{21}$\pcmsq. The lower flux spectrum, following the outflow phase reported here, indicates that there has been a three-fold reduction in \LX\ over the period of 12 hours (assuming a PO model), similar to the factor of $\sim$2 seen after three days in NGC 1365. 

\citet{Soria_07} suggest that such properties could be analogous to flares observed in Galactic X-ray binaries in a steep-power law state, where the outflow commences after the source exceeds the Eddington limit. Alternatively, some Galactic binaries have been observed to remain in a power-law spectrum HS during short outbursts \citep{Yu_09}. It has been suggested that some sources with such HS events could reach flux levels which classify them as ULX during their short outburst events \citep{Yu_09}, as we find in the case of A8. This further strengthens the suggestion of \citet{Yu_09}, that some ULXs which stay in the HS during a flaring event harbor stellar-mass compact stars. 
We note that Galactic BH binaries have been observed to typically proceed through hysteresis cycles \citep{Remillard_06} meaning that the HS event should be followed by a TD state, which is not observed in the subsequent pointing three months later. However, these canonical state transitions observed for Galactic binaries are often not observed in ULXs (e.g. \citealt{Feng_11} and references therein).

\subsection{Super Soft Sources}
\label{subsec:sss}

SSS are sources with little or no emission above 1 keV, with typical X-ray luminosities between 1$\times10^{36}$ to 1$\times10^{38}$ \ergps. Following the discovery of SSS in the Large Magellanic Cloud \citep{Long_81}, \citet{vandenh_92} proposed a model of quasi-steady nuclear burning on the surface of a white dwarf (WD) accreting matter from a Roche lobe-filling companion with high accretion rates, with emission described by a simple absorbed blackbody (BB). Further observations have revealed a heterogeneous class of objects with sources detected in both early- and late-type galaxies, in the field and in GCs, and with a range of temporal behavior, and luminosities as high as \LX$>$1$\times10^{39}$ \ergps, in the ULX range (e.g. \citealt{Carpano_07,Distefano_03,Fabbiano_03c}). It has been suggested that these sources could be WDs in super-Eddington outbursts, stellar-mass BHs, or even intermediate-mass black holes (e.g. \citealt{Distefano_04a}). 

A6 was observed in the second and third observations of NGC 3379 (Obs. ID 7073 \& 7074). The HR and colors do not change between pointings; consequently the spectra from these two observations were jointly fitted with all parameters apart from the normalization tied between observations. Of the three models, the PO does not provide a physically realistic description of the spectrum, determining a best-fit photon index of $\sim8$, and can be discounted. The best-fit parameters between the DBB and BB model are very similar, with an inner disc temperature of 91 $\pm$15 eV, and an absorption column consistent with the Galactic value for the BB model, and an \NH\ of more than twice the Galactic value for the DBB model, although within errors this is consistent with Galactic \NH. These best-fit models yield \LX=1.1$\times10^{38}$ \ergps\ (BB) and \LX=1.5$\times10^{38}$ \ergps\ (DBB), consistent within the uncertainties. 

The spectrum is that of a `classical' SSS. Due to its transient nature it is possible that the observed X-ray emission arises from a classical nova (CNe) where episodic thermonuclear explosions on the surface of the WD lasting from months to years can appear as a transient SSS \citep{Pietsch_07}. From X-ray/optical studies of M31 and M33 \citet{Pietsch_05} suggested that these sources are a major class of SSS.

\subsubsection{The extremely luminous SSS C1}
\label{subsec:c1sss}

\begin{figure*}
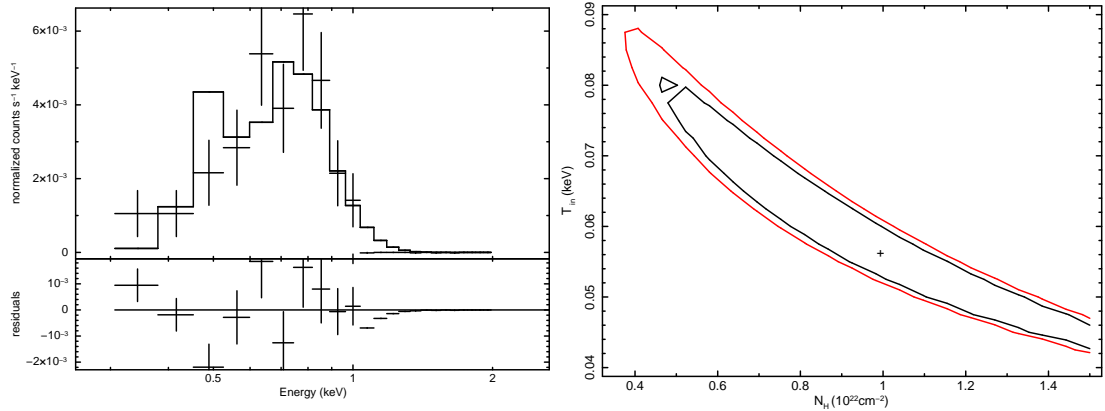

  \centering
	\begin{minipage}{0.4\linewidth}
	\centering

    \includegraphics[height=\linewidth,angle=-90]{f9a.ps}

	\end{minipage}
	\begin{minipage}{0.4\linewidth}
	\centering

    \includegraphics[height=\linewidth,angle=-90]{f9b.ps}

	\end{minipage}
	\centering

    \caption{Left: Spectrum and residuals of the SSS C1 from observation 4 (obs ID 4730) with a single-component BB model. Right: 1$\sigma$ and 2$\sigma$ contours plots of the BB model, the cross hair indicates the best-fit values. Details of this source are presented in Section \ref{subsec:c1sss}.  \label{fig:C1} }

\end{figure*}

The second SSS, C1, is even softer than A6, with an inner disc temperature of 60$\pm$10 eV and a large absorption column ($\sim 1\times10^{22}$\pcmsq), yielding an intrinsic extremely high X-ray luminosity of 5.8$\times10^{41}$ \ergps. The spectrum is shown in Figure \ref{fig:C1}, along with 1$\sigma$ and 2$\sigma$ contour parameters for the BB model. Even using the 1$\sigma$ lower limits of the inner disc temperature and absorption column of this fit,  the luminosity would still be extreme, 5.6$\times10^{40}$ \ergps. A neutron star atmosphere model applied to this spectrum also resulted in a large best-fit luminosity $>1~\times10^{41}$ \ergps. 

We further investigated if this source could be a foreground object (e.g. a flaring M dwarf) by fitting a thermal plasma model (APEC) to the spectrum. This resulted in a best-fit temperature $\sim$0.3 keV and a flux of $\sim1.1\times10^{-14}$ \flux\ (corresponding to \LX$\sim2\times10^{38}$ \ergps\ at the distance of NGC 4697). By assuming that C1 is a star it is possible to calculate the \LX/$L_{\rm{bol}}$ ratio by combining the photometric g-band upper limit for this source with the semi-empirical stellar SEDs of \citet{Kraus_07} to provide $L_{\rm{bol}}$, with \LX\ values scaled from the best-fit X-ray luminosity derived from the APEC model. The resulting value, \LX/$L_{\rm{bol}} >0.35$, is much higher than values that are typically seen in stars, where in a recent large survey \citep{Wright_11} \LX/$L_{\rm{bol}}$ values were 1$\times10^{-3}$ with a dispersion of an order of magnitude. Therefore, C1 is unlikely to be a foreground object unless it was observed during an incredibly rare flaring event for which the X-ray luminosity increased by more than an order of magnitude. To determine the likelihood of observing such a flare we compare to the studies of \citet{Feigelson_04} and \citet{Wright_10}, who both studied the properties of stellar X-ray sources in deep, high Galactic latitude fields. From a total of 71 sources and a cumulative exposure of 26.68 Ms they observed only 2 stars with flaring events with peak amplitudes greater than a factor of 10. From their detection thresholds and the relative exposure times we estimate that we should detect $\sim$5 stars in the observation of NGC 4697. This equates to a cumulative exposure of 660 ks on these stars, and therefore a probability of observing a significantly large flare in one of these stars of 0.05. This seems highly unlikely and therefore we conclude that the emission from C1 does not arise from a flare star.

C1 was previously discussed in S08 (Section 4.4), who, after determining that a background AGN is unlikely, suggest an intermediate mass black hole with a mass range of $\sim (7\times10^{3}-10^{6}) \Msol$, although they note that the formation of such a large black hole outside of the nucleus or a GC is a challenge for current black hole formation theory.

\begin{figure*}
  \centering
	\begin{minipage}{0.4\linewidth}
	\centering

    \includegraphics[height=\linewidth,angle=-90]{f10a.ps}

	\end{minipage}
	\begin{minipage}{0.4\linewidth}
	\centering

    \includegraphics[height=\linewidth,angle=-90]{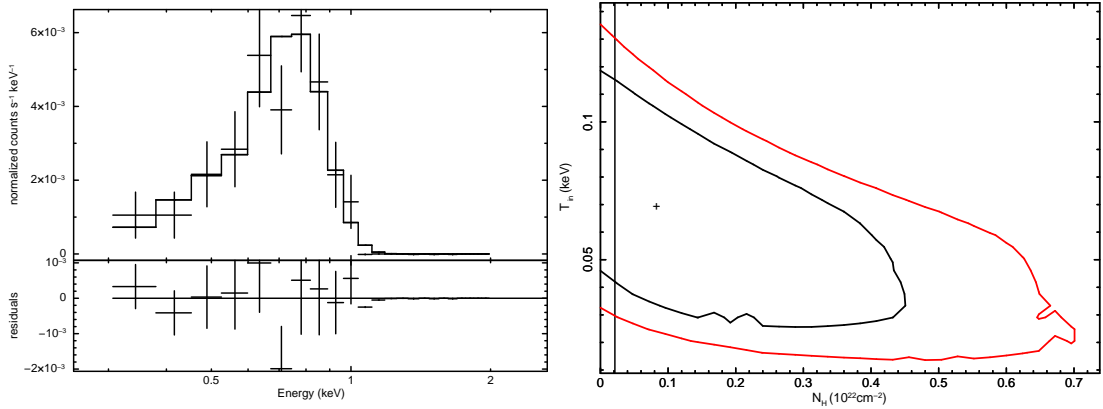}

	\end{minipage}
	\centering

    \caption{Left: Spectrum and residuals of the SSS C1 from observation 4 (obs ID 4730) fitted with a BB plus a Gaussian line model. Right: 1$\sigma$ and 2$\sigma$ contours plots of this BB+Gauss model, where the cross hair indicates the best-fit values. The vertical line denotes Galactic \NH. Details of C1 and the model presented here are discussed in Section \ref{subsec:c1sss}.  \label{fig:C1g} }

\end{figure*}

X-ray luminosities in excess of 10$^{41}$\ergps\ have only been reported for two previous SSS; ULX-1 in M101 \citep{Kong_05} and a luminous SSS transient in NGC 4631 \citep{Carpano_07}. Both of these galaxies are late-type star forming systems, unlike NGC 4697. These sources, when described by a single-component BB model, exhibited similar properties to C1, with soft X-ray spectra (\kt$\sim40-$150 eV), high intrinsic absorption, and resulting extremely large \LX. However, in both cases, alternate models have been suggested resulting in less extreme luminosities (\cite{Mukai_05} ; \citep{Liu_09}; \citet{Soria_09,Carpano_07}. For C1, even though a single-component model is statistically acceptable, there is a possible excess in the residuals around 0.8 keV (see Figure \ref{fig:C1}). Following \citet{Carpano_07} an additional Gaussian line was included with the BB model in the fit (we also attempted to use an absorption edge to model the spectrum, as presented in \citet{Soria_09} for the NGC 4631 source, but were able to constrain the model parameters). The values obtained from the BB+Gaussian model are presented in Table \ref{tab:C1} and the spectrum and the 1$\sigma$ and 2$\sigma$ contours are shown in Figure \ref{fig:C1g}. The best-fit parameters indicate a lower intrinsic absorption, consistent with the Galactic value within errors, and a temperature of 70 eV (compared to 60 eV in the single-component BB model). The peak energy of the Gaussian component, $\sim$0.7 keV, is likely to arise from a blend of unresolved emission lines (e.g. \OVIII; \citealt{Ness_05}), or there could alternatively be absorption edges, as seen in \citet{Soria_09}.
The resulting 0.3$-$8.0 keV luminosity, 3.4$\times10^{38}$ \ergps, although poorly constrained, is three orders of magnitude lower than that from the simple BB model. Although in excess of the emission from a hydrogen burning 1.35 \Msol\ WD system, this luminosity could arise from an extreme super Eddington event (a fireball scenario \citealt{Soria_09}).

The 2XMM survey reports a source within $<$2\arcsec of C1. Given its soft hardness ratios (e.g. HR2=-0.95$\pm0.08$\footnote{Derivation of Hardness Ratios are provided in section 3.1.2 in the \XMM\ Serendipitous Source Catalogue Users Guide}), and the lower spatial resolution of \XMM, it is possible that this is emission arising from C1. The \XMM\ observation was taken in July 2003 (compared to August 2004 for the `on' \CHANDRA\ observation) yielding an X-ray luminosity of 1.30$\times10^{38}$ \ergps, from the 2XMM catalogue (compared to 4.72$\times10^{38}$ \ergps\ for the model derived from the \CHANDRA\ observation over the 0.2$-$12 keV energy range). If this observed emission is from C1, it suggests that the source underwent two outbursts, with a recurrence time of $\sim$1 year. From simulations \citet{Starrfield_04} estimated that the hydrogen layer involved in the surface nuclear burning would have a mass of $<1\times10^{-6}$ \Msol\ and that steady surface hydrogen burning (for a 1.35\Msol\ WD) only occurs for accretion rates below 1$\times10^{-6}$ \Msolpy. If the outbursts of C1 are a consequence of this fireball scenario this implies accretion rates $>1\times10^{-6}$ \Msolpy\ and therefore the hydrogen shell could be replenished within 1 year. In fact, \citet{Soria_09} estimates that the SSS in NGC 4631 has an accretion rate of around 1$\times10^{-5}$ \Msolpy, which could therefore replenish the shell within a month.

However, this interpretation should be treated with caution due to the large uncertainties in the best-fit parameters. Furthermore, the inclusion of the Gaussian component, possibly related to the photospheric expansion period, is speculative, although it serves the purpose of lowering the intrinsic luminosity from the extreme value of $>5\times10^{41}$ \ergps\ to a more typical SSS value of 3.4$\times10^{38}$ \ergps.

\subsection{Quasi-Soft Sources}
\label{subsec:QSS}

QSS are systems with little or no emission above energies of 2 keV, with temperatures between 100 eV to 350 eV \citep{Distefano_04b}. QSS are too hot to be WDs, unless there is significant upscattering of photons emitted by the WD, or the emission emanates from a limited portion of the surface. Alternative scenarios include QSS as NS or BH binary systems \citep{Distefano_10} (and SNRs; \citealt{Orio_06}). The results of the spectral analysis put B8 and A5 in this category (see Table \ref{tab:Bestfit}). 

Source B8, was observed in one of the cycle 11 \CHANDRA\ observations and has a maximum outburst time of $\sim$3 years. Spectral modeling determines a temperature of between 280$-$360 eV and X-ray luminosity $\sim(1.1-1.5)\times10^{38}$\ergps\ (depending on if a DBB or BB model is preferred). Such values are consistent with the properties of previously observed QSS.

\subsubsection{The flaring QSS A5}
\label{subsec:srcFlare}

\begin{figure*}
  \centering

    \includegraphics[width=0.7\linewidth]{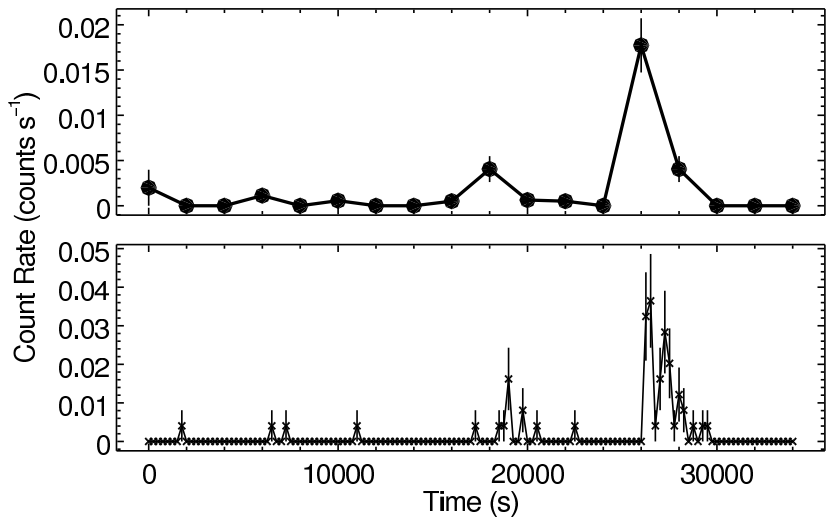}

    \caption{Short-term light curve of Source A5 from observation 1 (obs ID 1587). In the top panel binning of 2000s has been used and in the bottom panel a finer binning scheme of 250s is presented. The outburst lasts for less than 3500s. The unbinned light curve and event arrival times show that of the 61 events in this observation, 43 arise from a time interval $<$3260s. (Section \ref{subsec:srcFlare}). \label{fig:lc94} }

\end{figure*}

In the case of the flaring source A5, the best-fit DBB model yields inner disc temperature of 220 eV, with intrinsic absorption of $\sim$3$\times10^{21}$\pcmsq, Both the temperature and absorption column values from the BB model are lower than the DBB model, although are consistent within errors, and \LX=2.5 $\times10^{38}$\ergps. 
The outburst data (of $\sim$3 ks; Figure \ref{fig:lc94}) yield similar best-fit parameters to those determined from the full observation. However, due to the reduced statistics it was necessary to fix the absorption column; this was frozen to the Galactic value as well as the best-fit column density of 31$\times10^{21}$\pcmsq\ and 19$\times10^{21}$\pcmsq\ for the DBB and BB models respectively. 

Although the average best-fit luminosity of A5 is \LX=2.5 $\times10^{38}$\ergps, during the flare (see Table \ref{tab:Bestfit}), the luminosity reaches 2.9$\times10^{39}$ \ergps\ for the high absorption column, and 9.5$\times10^{38}$ \ergps\ for the Galactic absorption value for the DBB model; the corresponding values for the BB model are 1.7$\times10^{39}$ \ergps\ and 9.4$\times10^{38}$ \ergps.

This $\sim$3000s flaring event is not typical of QSS (e.g. \citealt{Distefano_04a,Liu_11}). Instead it could arise from a flare star and, to investigate this, a thermal plasma model was also fitted to the spectrum (e.g. \citealt{Wargelin_08}). This model resulted in a best-fit temperature of 1 keV and a flux of $\sim5\times10^{-14}$\flux\ (corresponding to \LX=$\sim1\times10^{39}$\ergps\ at the distance of NGC 3379). Using the g-band upper limit derived for A5 (see section \ref{sec:selection}), and following the same method described in section \ref{subsec:c1sss}, \LX/$L_{\rm{bol}} \sim1$. The most luminous stellar flares ever observed reach only \LX/$L_{\rm{bol}} \sim$0.1 \citep{Gudel_04} making this is an unfeasibly large value for a stellar source and making the possibility that this is a flaring star highly unlikely.

Instead this behavior is suggestive of a superburst from a NS. These events were first observed with BeppoSax \citep{Cornelisse_00} and exhibit outbursts of $\sim3000-10,000$s, with luminosities $\sim2-3\times10^{38}$\ergps\ and energetics $\sim1\times 10^{42}$ergs. Because of the large total energy released from these events, these bursts are thought to be powered by carbon burning instead of hydrogen and helium (see \citealt{Cumming_01,Cooper_05} and references therein), although it has also been suggested that this emission could arise from runaway electron capture \citep{Kuulkers_02}.

The spectra of superbursts are well described by a BB model, with temperatures $\sim$2 keV. For A5, when only the outburst spectrum is fitted, a temperature of $\sim200-300$ eV is determined. However, as we only model the high flux component of the spectrum, the luminosity of the source is \LX=$(1.7-2.9)\times10^{39}$\ergps\ when intrinsic absorption (constrained from the spectrum of the whole observation) is included. When \NH\ is fixed to the value of Galactic absorption \LX=$9.4\times10^{38}$\ergps, with a lower limit of $4\times10^{38}$\ergps. From these parameters the fluence of this source could be as high as 9.5$\times10^{42}$ergs (with a lower limit of 1.3$\times10^{42}$ergs). This emission is more luminous and energetic than the reported superburst sources, although both of these parameters are consistent within errors. However, the significant difference between A5 and superburst sources arises from the best-fit temperature, where A5 is much cooler. Under the assumption that the source is emitting isotropically a lower limit of R$>5\times10^7$ cm is derived. This value indicates that this source is larger than a NS and therefore cannot be a superburst.

Instead, given the implied radius (which has a maximum limit of R$<3.4\times10^8$ cm), this emission could arise from a WD. If such a high fluence event is powered by a thermo-nuclear processes this source could be a helium nova explosion. Such an object was first considered by \citet{Kato_89}, where the case of a degenerate WD accreting helium from its helium-rich companion was presented. In this scenario, for accretion rates $\sim\times10^{-9}-\times10^{-6}$\Msol\ yr$^{-1}$, when sufficient helium has been accumulated along a shell on the surface of the WD a thermo-nuclear reaction, or helium shell flash, occurs resulting in a nova outburst. In 2000 \citet{Kato_00} reported the discovery of such an object; V445 Puppis. This type of outburst event has also been predicted to emit in the X-ray, although possibly not until the post-outburst stage, before which the wind phase may lead to absorption of the soft X-rays \citep{Iben_94,Ashok_03}. However, the period of X-ray emission is expected to be similar to that of classical novae, which lasts for periods $>$months \citep{Pietsch_07}, therefore, an X-ray outburst of $\sim$3-ks indicates that this emission does not arise from a Helium Nova.

Another possibility is that this emission arises from a very short period system such as a double WD binary, where these short periods imply that mass transfer is driven by angular momentum loss via gravitational radiation. This would allow for the high X-ray luminosity observed and suggests a mass of 0.4$-$0.9\Msol\ for the accreting white dwarf in A5 (\LX=(4$-$9.4)$\times10^{38}$\ergps). Assuming that the mass transfer is driven by gravitational radiation gives donor masses slightly smaller than the accretors. Taking slightly smaller WD mass pairs gives a somewhat smaller average luminosity, allowing for some intrinsic variability, although the cause of this is unclear. A possible complication is that the Eddington luminosity may be depressed below the usual value for this type of accretion, since for CO-dominated compositions the Kramers opacity can exceed Thomson at high temperatures
\citep{Peacock_12}. Although `opacity' here refers to the Rosseland mean, and is therefore not strictly applicable for considering radiation pressure, a fuller exploration of this type of model would be needed.

\section{Conclusions}
\label{sec:conclusions}

In this paper we have identified the transient and potential transient candidates within three elliptical galaxies using deep multi-epoch \CHANDRA\ data. From spectral analysis of the seventeen sources that have been confirmed as a TC or PTCs (excluding the background source) a large variety of properties have been determined, revealing the heterogeneous nature of transient populations in elliptical galaxies. These sources are summarized in Table \ref{tab:summary}. 

From Table \ref{tab:summary} it can be seen that the majority of these sources have been revealed to be normally accreting LMXBs, where 12 of the objects have spectra consistent with binaries in either a hard or thermally dominant state. Of the remaining five sources, one (A8, which has also been determined to be a GC-LMXB) has been observed as a ULX, with a peak luminosity of 2.8$\times10^{39}$\ergps\ over a $\sim$12 hour period, after which the source drops to \LX=8$\times10^{38}$\ergps. During this bright period of emission the source has been determined to be in a hard state with evidence of enhanced ionized absorption with a column density of $\sim5\times10^{22}$\pcmsq\ and an ionization parameter of $\sim150$ \erg\ cm s$^{-1}$. Following this 12 hour period the lower flux emission is consistent with an LMXB in a typically HS. We suggest that this source has undergone a large flaring event prior to our observation and we have observed the subsequent period of decline, where an outflow phase has commenced after the super-Eddington accretion. This behavior is similar to the properties of the ULX observed in NGC 1365 \citep{Soria_07} and could indicate that A8 is a stellar-mass BH binary undergoing a HS flaring event.

The two SSS transients presented in this paper have properties indicating that they are both likely to arise from WD binaries. The first, A6, has spectra consistent with `typical' SSS, with a temperature of 90 eV and, given the transient nature of this source, is likely to be a classical nova. The second SSS, when fitted with a single-component DBB model has \LX$>5\times10^{41}$\ergps. This value is much greater than any of the previously confirmed X-ray luminosities of ultra-luminous SSS and is difficult to explain. Instead, we include a Gaussian component to the model, which we suggest could be related to an outflow or expansion, as seen in NGC 4361 \citep{Soria_09}. Our tentative interpretation of this source is a WD binary with photospheric expansion taking place, resulting in an extreme super Eddington event with an X-ray luminosity $\sim3.4\times10^{38}$\ergps. 

In addition to the SSS, two of the transients presented here have been classified as QSS. One of these sources (B8) has spectral properties that are consistent with previously observed QSS, with a temperature $\sim$300 eV and a luminosity $\sim10^{38}$ \ergps. Source A5 on the other hand has rather unusual temporal behavior, where $>$70\% of the emission from this source arises within 3 ks, with a peak luminosity of 9.4$\times10^{38}$\ergps\ (with a lower limit of 4$\times10^{38}$\ergps) and a temperature of $\sim200-300$ eV. This cool emission indicates that the source is a WD and, as a consequence of this short-lived highly luminous event, we tentatively suggest that this emission arises from super-Eddington accretion from a short period double WD binary.

From cross-correlations of {\em HST} data (presented in B08, B09, S08 and section \ref{sec:selection} of this paper) we have been able to quantify the number of field and GC transients and determine the nature of binaries in these objects. Only three sources (A2, A8 \& B1) have been determined to be GC-LMXBs. From the modeling of B1 this source could be emitting in either a hard or thermally dominant state but, due to the high luminosity of $>5\times10^{38}$\ergps, we favor the interpretation of thermal emission, providing a best-fit inner disc temperature of 1.3 keV. 
Due to the high luminosity of not only this object, which is above the Eddington limit for a NS, but also A8, both sources are BH-GC candidates. Such objects are expected to be rare and these provide only the sixth and seventh unambiguous example of such a system. Further, these are only the second and third BH-GC that has been shown to exhibit transient behavior (although a BH-GC candidate emitting $\sim L_{\rm{Edd}}$ has recently been presented in \citealt{Barnard_11}). Due to the transient nature of these systems we suggest a formation channel of an exchange interaction \citep{Kalogera_04}, instead of a tidally captured source (or an ultra-compact binary; \citealt{Gnedin_09,King_11}) which would lead to the formation of a persistent BH-GC.

From the optical matching we have determined that 14 sources arise from the field compared to only three from GCs (the remaining source is a background object). The large number of field sources in the transient population is expected as theoretical work predicts that the majority of the field population will arise from relatively detached systems, which will be transient $>$75\% of their lifetime. 

From population synthesis modeling \citep{Fragos_09} predictions were made about the number of TCs and PTCs that would be detected in our new observations of NGC 4278. However, from comparing the newly determined numbers of TC/PTC sources to those predicted we have much fewer sources, in fact no new TCs were observed and only 2 PTCs, compared to the prediction of 6 TCs and 6 PTCs. We attribute this discrepancy to the original comparison of the modeling results to the observed transient population, where all transient sources included in B08 and B09 were assumed to be NS-LMXBs, and significant difference in our and their statistical analysis and error estimates.


\acknowledgments

 We thank the CXC DS and SDS teams for their efforts in reducing the data and 
developing the software used for the reduction (SDP) and analysis
(CIAO). We would also like to thank the anonymous referee whose detailed and careful report has helped to improve this paper.
This work was supported by {\em Chandra} G0 grant G06-7079A \& GO0-11102X (PI:Fabbiano). We acknowledge partial support from NASA contract NAS8-39073(CXC). A. Kundu acknowledges support for this work by NASA through Chandra awards GO-11111 and AR-12009. A. Zezas acknowledges support from NASA LTSA grant NAG5-13056. S. Pellegrini and G. Trinchieri acknowledge partial financial support from the Italian Space Agency ASI (Agenzia Spaziale Italiana) through grant ASI-INAF I/009/10/0. S. Zepf acknowledges support from NASA ADAP grants NNX08AJ60G and NNX11AG12G. G. Fabbiano is grateful to the Aspen Center for Physics for their hospitality


\bibliography{nicky}

\clearpage

\LongTables

\begin{deluxetable}{cccccccccc}
\tabletypesize{\scriptsize}
\tablecolumns{5}
\tablewidth{0pt}
\tablecaption{Summary of the properties of the three galaxies \label{tab:prop}}
\tablehead{\colhead{Galaxy} & \colhead{$D$} & \colhead{\LB} &
\colhead{Age} & \colhead{$S_{\mathrm{GC}}$} & \colhead{Obs ID} & \colhead{Obs Date}  & \colhead{Expo} & \colhead{No. Src} \\
\colhead{(1)} & \colhead{(2)} & \colhead{(3)} &
\colhead{(4)} & \colhead{(5)} & \colhead{(6)} & \colhead{(7)}  & \colhead{(8)} & \colhead{(9)} }
\startdata
NGC 3379  & 10.6 & 1.46  &    9.3$-$14.7$^{a,b}$  &  1.2 &      &          &      &    \\
          &      &       &                &      & 1587 & Feb 2001 & 29.0 & 58 \\
          &      &       &                &      & 7073 & Jan 2006 & 80.3 & 72 \\
          &      &       &                &      & 7074 & Apr 2006 & 66.7 & 69 \\
          &      &       &                &      & 7075 & Jul 2006 & 79.6 & 77 \\
          &      &       &                &      & 7076 & Jan 2007 & 68.7 & 63 \\
          &      &       &                &      & All  &      -   &324.2 & 125 (+7) \\
NGC 4278  & 16.1 & 1.63  &    10.7$-$14.1$^{a,b}$  & 6.9 &      &          &      &    \\
          &      &       &                &      & 4741 & Feb 2005 & 37.3 & 85 \\
          &      &       &                &      & 7077 & Mar 2006 & 107.7 & 152 \\
          &      &       &                &      & 7078 & Jul 2006 & 48.1 & 84 \\
          &      &       &                &      & 7079 & Oct 2006 & 102.5 & 122 \\
          &      &       &                &      & 7081 & Feb 2007 & 54.8 & 141 \\
          &      &       &                &      & 7080 & Apr 2007 & 107.6 & 105 \\
          &      &       &                &      & 11269 & Mar 2010 & 80.7 & 116 \\
          &      &       &                &      & 12124 & Mar 2010 & 25.5 & 64 \\
          &      &       &                &      & All  &      -   & 564.2 & 231 (+20) \\
NGC 4697  & 11.8 & 2.00  &    8.1$-$11.0$^{a,c}$  & 2.5   &      &          &      &   \\
          &      &       &                &      & 0784 & Jan 2000 & 36.7 & 95 \\
          &      &       &                &      & 4727 & Dec 2003 & 36.6 & 72 \\
          &      &       &                &      & 4728 & Jan 2004 & 33.3 & 74 \\
          &      &       &                &      & 4729 & Feb 2004 & 22.3 & 59 \\
          &      &       &                &      & 4730 & Aug 2004 & 38.1 & 82 \\
          &      &       &                &      & All  &      -   & 132.0 & 120 (+4) \\
\enddata
\tablecomments{No. src denotes the number of sources detected within each observation in the overlapping region of the S3 chip in all observations. Also note there that in NGC 4278 obs IDs 7081 was taken prior to obs ID 7080, we therefore retain the chronological sequence with obs ID 7081 labeled as pointing 5 in subsequent figures. Age estimates from ($a$) \citet{Terlevich_02}, ($b$) \citet{Kuntschner_10}, ($c$) \citet{Rogers_10}. }
\end{deluxetable}


\begin{deluxetable}{c@{}c@{}c@{}c@{}c@{}c@{}c@{}c@{}c@{}c@{}c@{}c@{}}
\tabletypesize{\scriptsize}
\tablecolumns{12}
\tablewidth{0pt}
\tablecaption{Properties of TCs and PTCs detected in the three galaxies \label{tab:tcs}}
\tablehead{\colhead{Src No.} & \colhead{Catalog No.} & \colhead{RA} & \colhead{Dec} & \colhead{OBS ID} &
\colhead{Net Counts} & \colhead{Mode Ratio} & \colhead{Lower-bound value} & \colhead{TC/PTC} & \colhead{Light Curve} & \colhead{Notes} & \colhead{Optical Limits} \\
\colhead{(1)} & \colhead{(2)} & \colhead{(3)} &
\colhead{(4)} & \colhead{(5)} & \colhead{(6)} & \colhead{(7)}  & \colhead{(8)} & \colhead{(9)} & \colhead{(10)} & \colhead{(11)} & \colhead{(12)} }
\startdata
A1 & 25 (B08)  & 10:47:54.5 & +12:35:30.9 & 7074  & 174.9  & 161.4 & 43.5 & TC  & Mid (2)  & Field & $>$27.4 \\
-  & -         & -          & -           & 7075  & 24.2   & -     & -    & -   & -        &   -   &   -  \\
A2 & 50 (B08)  & 10:47:51.6 & +12:35:35.9 & 7073  & 44.1   & 6.8   & 5.9  & PTC & LMid (1) & GC    &   -  \\
A3 & 85 (B08)  & 10:47:49.2 & +12:34:31.1 & 7074  & 38.4   & 30.2  & 9.0  & PTC & Mid (1)  & Background &   -   \\
A4 & 89 (B08)  & 10:47:48.9 & +12:34:59.1 & 7076  & 97.6   & 18.4  & 16.4 & TC  & Beg (1)  & Field & $>$25.0  \\
A5 & 94 (B08)  & 10:47:48.0 & +12:34:45.7 & 1587  & 58.3   & 19.2  & 13.2 & TC  & End (1)  & Field/Double &$>$26.2 \\
A6 & 100 (B08) & 10:47:47.2 & +12:34:59.8 & 7073  & 54.8   & 20.2  & 14.1 & TC  & LMid (2) & Field &$>$26.4 \\
-  &  -        & -          & -           & 7074  & 19.5   &    -  & -    & -   & -        &   -  &   -  \\
A7 & 115 (B08) & 10:47:43.6 & +12:34:10.8 & 1587  & 18.3   & 12.9  & 6.8  & PTC & End (1)  & Field &$>$27.0 \\
A8 & 128 (B08) & 10:47:38.2 & +12:33:07.6 & 7073  & 548.3  & 85.1  & 50.7 & TC  & LMid (2) & GC &   - \\
-  &    -      & -          & -           & 7074  & 9.1    &   -   &  -   &  -  & -        &   -  &   - \\
\\
\hline
\\
B1 & 83 (B09)  & 12:20:08.138 & +29:17:17.1 & 7079  & 249.7  & 57.8  & 35.5  & TC & Mid (1)  &  GC  &   -  \\
B2 & 91 (B09)  & 12:20:07.930 & +29:16:57.2 & 7079  & 93.5   & 9.0   & 8.3   & PTC & Mid (1) & Field & $>$26.2  \\
B3 & 139 (B09) & 12:20:06.218 & +29:16:57.5 & 7079  & 170.8  & 26.1  & 18.9  & TC & Mid (1)  & Field & $>$25.8 \\
B4 & 156 (B09) & 12:20:05.669 & +29:17:15.9 & 7077  & 50.5   &  6.2  & 5.5   & PTC & LMid (1) & Field & $>$26.5  \\
B5 & 159 (B09) & 12:20:05.388 & +29:17:17.0 & 7080  & 23.2   & 13.6  & 5.8   & PTC & LMid (1)  & Field & $>$26.6  \\
B6 & 204 (B09) & 12:20:01.582 & +29:16:26.2 & 4741  & 339.4  & 871.7 & 202.9 & TC & End (1) & Field & $>$27.0  \\
B7 & - (new)   & 12:20:09.001 & +29:17:23.7 & 11269 & 70.6   & 11.2  & 8.4   & PTC & Mid (2) & Field & $>$27.0  \\
-  &    -      &  -           & -           &12124  & 16.1  &   -   &  -    &  -  & -        &   -   &   -  \\
B8 & - (new)   & 12:20:06.361 & +29:16:54.6 & 11269 & 66.1   & 11.6  & 8.7   & PTC & Beg (1) & Field & $>$24.9  \\
\\
\hline
\\
C1 & 110 (S08)  & 12:48:40.507 & -05:50:20.7 & 4730  & 87.4  & 30.9 & 22.9 & TC & Beg (1)  &  Field & $>$27.0 \\
C2 & 104 (S08)  & 12:48:27.484 & -05:49:00.8 & 4730  & 16.5  & 8.0 & 5.4 & PTC  & Beg (1)  &  Field & $>$27.8 \\
\enddata
\tablecomments{All net count values are derived from the standard processing techniques detailed in both B08 and B09. Optical limits are provided for sources without an identified optical counterpart, where 3$\sigma$ g-band (Vega) magnitudes are quoted.  In column 10 sources that could be in outburst for over a year are labeled as `LMid' and those with a shorter outburst upperlimit are `Mid'. Sources that were observed in the first observation are labeled as `End' and sources labeled as `Beg' are seen at the beginning of their `on state' in the last epoch of each galaxy's monitoring campaign. The number in brackets in this column indicates the number of observations the source was detected in.}
\end{deluxetable}

\clearpage

\begin{deluxetable}{ccccc}
\tabletypesize{\scriptsize}
\tablecolumns{5}
\tablewidth{0pt}
\tablecaption{Hardness ratios and colors of TCs and PTCs detected in the three galaxies \label{tab:phot}}
\tablehead{\colhead{Src No.} & \colhead{HR} & \colhead{C21} &
\colhead{C32} & \colhead{Log \LX} \\
\colhead{} & \colhead{} & \colhead{} &
\colhead{} & \colhead{erg $s^{-1}$} \\
\colhead{(1)} & \colhead{(2)} & \colhead{(3)} &
\colhead{(4)} & \colhead{(5)} }
\startdata
A1 & -0.73    $_ { -0.06}^ { +0.04}$  & -0.16$_ { -0.10}^ { +0.05}$    &  0.76$_ { -0.10}^ { +0.13}$& 38.4 \\
A2 & -0.20$^{+0.14}_{-0.16}$ & -0.48$_ { -0.22}^ { +0.19}$ & 0.43$_ { -0.16}^ { +0.18}$  & 37.8 \\
A3 & -0.46     $_ { -0.16}^ { +0.15}$  &  -0.04$_ { -0.18}^ { +0.16}$   &       0.22$_ { -0.18}^ { +0.19}$& 37.8 \\
A4 & -0.42   $_ { -0.11}^ { +0.09}$  &  -0.11$_ { -0.12}^ { +0.10}$     &  0.32$_ { -0.09}^ { +0.14}$  & 38.1 \\
A5 & -0.92       $_ { -0.06}^ { +0.04}$  &       -0.14$_ { -0.11}^ { +0.13}$ &  1.16$_ { -0.35}^ { +0.58}$  & 38.3 \\
A6 & -1.00    $_ { -0.00}^ { +0.02}$  &       1.99$_ { -0.54}^ { +0.91}$      &       -0.08$_ { -1.14}^ { +1.15}$ & 37.8 \\
A7 & -0.65        $_ { -0.21}^ { +0.17}$  &       0.00$_ { -0.21}^ { +0.21}$      &       0.64$_ { -0.40}^ { +0.57}$ & 37.8  \\
A8 & -0.28      $_ { -0.04}^ { +0.04}$  &       -0.27$_ { -0.02}^ { +0.08}$     &       0.22$_ { -0.04}^ { +0.05}$ & 38.9 \\
\\
\hline
\\
B1 & -0.52$_{ -0.06}^ { +0.05}$ & -0.29$_ { -0.03}^ { +0.10}$     &   0.58$_ { -0.09}^ { +0.07}$ & 38.7 \\
B2 & -0.83$_{ -0.07}^ { +0.05}$ & 0.04$_ { -0.11}^ { +0.08}$      &   1.08$_ { -0.25}^ { +0.43}$ & 38.2 \\
B3 & -0.79$_{-0.05}^ { +0.06}$  & -0.23$_ { -0.09}^ { +0.08}$     &  1.00$_ { -0.16}^ { +0.20}$  & 38.5 \\
B4 & -0.58$_{ -0.14}^{+0.12}$   & -0.37$_ { -0.17}^ { +0.16}$     &  0.80$_ { -0.24}^ { +0.41}$  & 38.0 \\
B5 & -0.08$_ { -0.25}^{ +0.24}$ &-0.24$_ { -0.32}^ { +0.24}$     &   0.16$_ { -0.24}^ { +0.24}$  & 37.9 \\
B6 & -0.61$_ { -0.05}^{ +0.04}$ &-0.19$_ { -0.07}^ { +0.03}$     &   0.68$_ { -0.08}^ { +0.06}$  & 39.2 \\
B7 & -0.31$_ { -0.13}^{ +0.10}$ & -0.15$_{ -0.17}^ { +0.12}$     & 0.18$_ { -0.15}^ { +0.09}$    & 38.2  \\
B8 & -1.00$_ { -0.00}^{ +0.02}$ & -0.06$_{ -0.12}^ { +0.16}$      & 1.66$_ { -0.68}^ { +0.67}$   & 38.2   \\
\\
\hline
\\
C1 & -0.98$_{ -0.02}^ { +0.02}$ & 1.09$_ { -0.13}^ { +0.19}$     &   0.59$_ { -0.40}^ { +0.54}$ & 38.4 \\
C2 & -0.20$_{ -0.27}^ { +0.25}$ & 0.03$_ { -0.24}^ { +0.24}$     &   0.35$_ { -0.32}^ { +0.40}$ & 37.7 \\
\enddata
\tablecomments{The definition of these values are explained in Sections 2.1 \& 2.3 of B08.}
\end{deluxetable}

\clearpage

\begin{deluxetable}{ccccccccccc}
\tabletypesize{\scriptsize}
\tablecolumns{11}
\tablewidth{0pt}
\tablecaption{Summary of the best-fit parameters of the source spectra. Errors are given to 1$\sigma$ \label{tab:Bestfit}}
\tablehead{ \colhead{Source} & \colhead{Obs. ID.} & \colhead{Net Counts} & \colhead{Model} & \colhead{$\chi^2$/$\nu$} &
\colhead{$P_{\chi^2}$} & \multicolumn{3}{c}{Parameters} & \colhead{\ensuremath{L_{\mathrm{X}}}} & \colhead{\ensuremath{L_{\mathrm{X}}} Range}  \\
\colhead{}   &
\colhead{}  & \colhead{} & \colhead{} & \colhead{}  & \colhead{} &
\colhead{\ensuremath{N_{\mathrm{H}}}} & \colhead{$\Gamma$}  & \colhead{\kt} &
\colhead{$\times 10^{38}$ erg $s^{-1}$} & \colhead{}  \\
\colhead{}  & \colhead{}  &
\colhead{}  & \colhead{} & \colhead{}  & \colhead{} &
\colhead{$\times 10^{20}$} & \colhead{}  & \colhead{} &
\colhead{(0.3$-$8.0 keV)} \\
\colhead{(1)} & \colhead{(2)} & \colhead{(3)} &
\colhead{(4)} & \colhead{(5)} & \colhead{(6)} & \colhead{(7)}  & \colhead{(8)} & \colhead{(9)} & \colhead{(10)} & \colhead{(11)}}
\startdata
A1 & 7074	& 153.5 & Power Law	& 238C       & 44\%C	& 20.3$^{+8.3}_{-7.4}$	        & 2.59$^{+0.33}_{-0.31}$	& -	                        & 4.89 	& 4.12$-$5.39 \\
	&	&       &  DBB		& 236C      & 93\%C	& 2.8F				& -				& 0.64$^{+0.08}_{-0.07}$	& 2.42	& 1.48$-$2.62 \\
\\
A2 & 7073  & 42.2  & Power Law     & 191C      & 77\%C    & 2.8F                             & 1.70F                         & -                             & 0.79  & 0.48$-$1.06 \\
        &       &       & DBB           & 188C      & 29\%C    & 2.8F                             & -                             & 1.00F                         & 0.62  & 0.41$-$0.83  \\
\\
A3 & 7074  & 31.4  & Power Law     & 148C      & 54\%C    & 2.8F                             & 1.70F                         & -                             & 0.66  & 0.43$-$1.00  \\
        &       &       & DBB           & 149C      & 67\%C    & 2.8F                             & -                             & 1.0F                          & 0.53  & 0.27$-$0.75  \\
\\
A4 & 7076  & 87.3  & Power Law     & 247C      & 77\%C    & 2.8F                             & 1.64$^{+0.19}_{-0.19}$        & -                             & 2.14  & 1.68$-$2.52  \\
        &       &       & DBB           & 247C      & 85\%C    & 2.8F                             & -                             & 1.16$^{+0.26}_{-0.22}$        & 1.74  & 1.08$-$1.98  \\
\\
A5 & 1587  & 60.0  & Power Law     & 111C      & 95\%C    & 75$^{+31}_{-22}$                & 6.1$^{+1.7}_{-1.0}$        & -                             & 139  & 108$-$170  \\
        &       &       & DBB           & 108C      & 82\%C    & 30$^{+18}_{-14}$           & -                             & 0.22$^{+0.06}_{-0.04}$        & 4.42  & 0.81$-$5.42  \\
        &       &       & BB           & 108C      & 60\%C    & 19$^{+19}_{-15}$           & -                             & 0.19$^{+0.04}_{-0.04}$        & 2.46  & 0.00$-$2.89  \\
\\
Flare   & 1587  & 41.9  & Power Law     & 114C       & 54\%C    & 2.8F                          & 2.42$^{+0.29}_{-0.26}$        & -                             & 14.11 & 10.46$-$25.72  \\
        &       &       & DBB           & 92C        & 69\%C    & 30.7F                         & -                             & 0.22 $^{+0.03}_{-0.02}$       & 29.11 & 20.10$-$33.18  \\
        &       &       & DBB           & 99C        & 16\%C    & 2.8F                          & -                             & 0.33$^{+0.06}_{-0.05}$        & 9.46  & 4.15$-$10.01   \\
        &       &       & BB           & 90C        & 89\%C    & 19.2F                         & -                             & 0.19$^{+0.02}_{-0.02}$        & 16.50 & 10.41$-$20.00   \\
        &       &       & BB           & 94C        & 73\%C    & 2.8F                         & -                             & 0.22$^{+0.02}_{-0.03}$        & 9.42  & 5.69$-$12.72   \\
\\
A6 & 7073	& 78.3	& Power Law	& 43C	     & 12\%C	& 20$^{+7}_{-6}$	        & 8.19$^{+0.94}_{-0.73}$	& -	                        & 6.26  & 1.91$-$10.12 \\
        & \& 7074	&	&  DBB		& 87C      & 11\%C	& 8$^{+9}_{-6}$		& -				& 0.09$^{+0.02}_{-0.01}$	& 1.54	& 0.54$-$2.82 \\
        & 	&	&  BB		& 94C      & 2\%C	& 2$^{+5}_{-2}$		& -				& 0.09$^{+0.02}_{-0.01}$	& 1.04	& 0.00$-$8.31 \\
        & 	&	&  BB		& 94C      & 4\%C	& 2.8F          		& -				& 0.09$^{+0.02}_{-0.01}$	& 1.12	& 0.622$-$1.67 \\
\\
A7  & 1587  & 17.8  & Power Law	& 75C      & 67\%C	& 2.8F              		& 1.70F                 	& -                             & 0.75  & 0.49$-$1.20 \\
         &       &       & DBB          & 77C      & 79\%C	& 2.8F                          & -                             & 1.00F                         & 0.61  & 0.34$-$0.92  \\
\\
A8 & 7073	& 521.2	& Power Law	& 20.5/23	& 0.61	& 2.8F				& 1.11$^{+0.08}_{-0.07}$	& -	                        & 14.83	& 13.87$-$16.23 \\
        & 	&	& DBB		& 27.2/23	& 0.25	& 2.8F				& -				& 2.60$^{+0.34}_{-0.46}$	& 13.97	& 5.01$-$15.52 \\
\hline
B1 & 7079	& 222.2 & Power Law	& 2.80/8	& 0.09	& 1.1$^{+5.7}_{-1.1}$		& 1.59$^{+0.20}_{-0.14}$	& -                             & 6.33  & 5.90$-$7.46 \\
	&	&	& Power Law	& 3.33/9	& 0.10	& 1.8F                 		& 1.60$^{+0.16}_{-0.15}$	& -                             & 6.34	& 5.63$-$7.43 \\
        &       &       & DBB           & 6.45/9	& 0.07	& 1.8F				& -                             & 1.28$^{+0.12}_{-0.23}$        & 5.36  & 2.97$-$5.97  \\
\\
B2 & 7079  & 100.1  & Power Law	& 184C	        & 66\%C	& 16.6$^{+12.2}_{-9.2}$		& 3.25$^{+0.72}_{-0.56}$	& -                             & 1.87  & 1.81$-$2.33 \\
        &       &       & DBB           & 181C		& 72\%C	& 1.8F				& -                             & 0.41$^{+0.07}_{-0.05}$        & 1.03  & 0.66$-$1.07  \\
\\
B3 & 7079  & 163.1 & Power Law	& 224C	        & 66\%C	& 29.8$^{+9.6}_{-8.8}$		& 3.21$^{+0.50}_{-0.45}$	& -                             & 9.57  & 7.33$-$9.95 \\
        &       &       & DBB           & 220C		& 60\%C	& 4.9$^{+4.7}_{-4.9}$   	& -                             & 0.55$^{+0.13}_{-0.07}$        & 2.57  & 1.67$-$2.77  \\
\\
B4 & 7077  & 53.1	& Power Law & 173C	& 64\%C	& 1.8F		                & 1.7$^{+0.23}_{-0.22}$	        & -                             & 1.22  & 0.90$-$1.64 \\
	&	&	& DBB		    & 166C	& 71\%C	& 1.8F				& -				& 0.79$^{+0.18}_{-0.12}$        & 0.84  & 0.42$-$1.05 \\
\\
B5 & 7080  & 20.5	& Power Law	& 114C	& 76\%C	& 1.8F			        & 1.7F				& -                             & 0.63  & 0.38$-$1.17 \\
	&	&	        & DBB		& 112C	& 52\%C	& 1.8F				& -				& 1.0F				& 0.53  & 0.23$-$0.91 \\
\\
B6 & 4741  & 313.1	& Power Law	& 12.60/12	& 0.40	& 9.4$^{+4.1}_{-5.1}$	& 2.00$^{+0.14}_{-0.18}$	& -                             & 23.22 & 22.28$-$25.91 \\
	&	&	& DBB		& 18.5/13	& 0.14	& 1.8F				& -				& 0.82$^{+0.09}_{-0.09}$	& 14.62 & 10.13$-$15.19 \\
\\
B7 	& 11269 & 87.1	& Power Law & 317C	& 95\%C	& 1.8F		                & 1.3$^{+0.35}_{-0.30}$	        & -                             & 2.53  & 1.86$-$3.80 \\
	& \& 12124	&	& DBB  & 317C	& 97\%C	& 1.8F				& -				& 1.16$^{+1.01}_{-0.34}$        & 1.63  & 1.13$-$2.13 \\
\\
B8 	& 11269 & 60.38	& Power Law	& 143C	& 24\%C	& 1.8F			        & 2.40$^{+0.33}_{-0.20}$		& -                             & 1.73  & 1.39$-$2.36 \\
	&	&	& DBB		& 131C	& 55\%C	& 8.8$^{+10.3}_{-8.6}$		& -				& 0.36$^{+0.09}_{-0.07}$   	& 1.55  & 0.54$-$1.64 \\
	&	&	& BB		& 131C	& 67\%C	& 1.8F             		& -				& 0.28$^{+0.03}_{-0.03}$   	& 1.11  & 0.82$-$1.39 \\
\hline
C1 & 4730  & 86.9  & Power Law	& 100C		& 50\%C	& 82.2$^{+3.6}_{-22.1}$		& 9.87$^{+0.03}_{-9.87}$	& -                             & 11590 & unconstrained \\
        &       &       & DBB           & 69C		& 57\%C	& 110$^{+44}_{-31}$		& -                             & 0.06$^{+0.02}_{-0.01}$        & 16557  & 8160$-$18670 \\
        &       &       & BB            & 67C   	& 58\%C	& 100$^{+32}_{-31}$  	        & -                             & 0.06$^{+0.01}_{-0.01}$        & 5752  & 4400$-$7076   \\

\\
C2 & 4730	& 15.8	& Power Law	& 91C	        & 86\%C	& 2.1F			        & 1.7F				& -				& 0.40	& 0.17$-$0.69 \\
	&	&	& DBB		& 92C		& 83\%C	& 2.1F				& -				& 1.00F				& 0.31	& 0.15$-$0.58 
\enddata
\tablecomments{These spectral models are presented and discussed in Sections \ref{subsec:specanaly} \& \ref{sec:results}.}
\end{deluxetable}


\begin{deluxetable}{cccc}
\tabletypesize{\scriptsize}
\tablecolumns{4}
\tablewidth{0pt}
\tablecaption{Best-fit parameters of A8 for the whole spectrum and the spectra from the high and low flux states. Presented are values from the tbabs * tbabs * absori * power law model with the quoted errors given to 1$\sigma$. \label{tab:spec128}}
\tablehead{\colhead{Parameters} & \colhead{Whole Spectrum} & \colhead{High Flux Spectrum} &  \colhead{Low Flux Spectrum} \\
 \colhead{(1)} & \colhead{(2)} & \colhead{(3)} & \colhead{(4)}  }
\startdata

Net Counts                             & 521.2                  & 366                      & 130   \\
$\chi^2$/$\nu$                         &         13.8/20        & 13.2/13                  & 303C  \\
$P_{\chi^2}$                          & 0.84                   & 0.04                     & 44\%C \\
\NH\ (10$^{21}$ \pcmsq )                & 0.43$^{+0.54}_{-0.43}$ &  1.79$^{+0.99}_{-1.26}$  & 0.34$^{+0.68}_{-0.39}$ \\
$\Gamma$                               & 1.45$^{+0.17}_{-0.23}$ & 1.89$^{+0.35}_{-0.41}$   &  1.04$^{+0.24}_{-0.22}$ \\
$k_{\mathrm{po}} (10^{-5}$ )           &  1.2$^{+0.8}_{-0.4}$   & 3.7$^{+3.1}_{-1.7}$      &  - \\
$T^{\mathrm{abs}}$ (K)                 & $10^{5}$F              & $10^{5}$F                &  - \\
$Z^{\mathrm{abs}}$ (\Zsol)             & 1F                     & 1F                       &  - \\
$\xi^{\mathrm{abs}}$ (\erg\ cm s$^{-1}$)             &  133$^{+113}_{-39}$     & 149$^{+79}_{-52}$       &  - \\
\NH$^{\mathrm{abs}}$  (10$^{22}$ \pcmsq ) & 2.33$^{+2.23}_{-1.46}$ &  4.95$^{+2.84}_{-2.25}$ &  - \\
\LX\ (10$^{39}$ \ergps )                &  1.59                   & 2.75                    &  0.89        \\
\LX\ Range  (10$^{39}$ \ergps )                              & 1.39$-$1.93             & 1.97$-$3.14             &  0.80$-$1.03 \\
\enddata
\tablecomments{The difference in counts between the `whole' observation and the high and low flux spectra arises from the short time gap between the two extracted spectra. This model is discussed in detail in Section \ref{subsec:srcA8}.}
\end{deluxetable}

\begin{deluxetable}{cc}
\tabletypesize{\scriptsize}
\tablecolumns{2}
\tablewidth{0pt}
\tablecaption{Best fitting parameters of SSS C1. Presented are values from the tbabs * BB +Gauss model with the quoted errors given to 1$\sigma$. \label{tab:C1}}
\tablehead{\colhead{Parameters} & \colhead{Best-fit Values} \\
 \colhead{(1)} & \colhead{(2)}  }
\startdata

Cstat                          & 52C   \\
Goodness                       & 10\% \\
\NH\ (10$^{21}$ \pcmsq )         & 0.85$^{+0.31}_{-0.85}$ \\
\kt\ (eV)                       & 70$^{+33}_{-49}$  \\
$E_{\rm{L}}$ (keV)              & 0.71$^{+0.04}_{-0.09}$ \\
$\sigma_{\rm{L}}$ (keV)         & 0.12$^{+0.04}_{-0.01}$  \\
\LX\ (10$^{38}$ \ergps )        & 3.40 \\
\LX\ Range  (10$^{38}$ \ergps)   & 0.00$-$3.90 \\
\enddata
\tablecomments{This spectral model and subsequent interpretation of the source is presented in Section \ref{subsec:c1sss}.}
\end{deluxetable}

\begin{deluxetable}{ccccc}
\tabletypesize{\scriptsize}
\tablecolumns{5}
\tablewidth{0pt}
\tablecaption{Summary of the properties of the seventeen TC/PTC sources \label{tab:summary}}
\tablehead{\colhead{Src No.} & \colhead{Optical Correlation} & \colhead{Spectral State} &
\colhead{\LX} & \colhead{Suggested Nature} \\
\colhead{} & \colhead{} & \colhead{} &
\colhead{$\times 10^{38}$ erg $s^{-1}$} & \colhead{} \\
colhead{} & \colhead{} & \colhead{} &
\colhead{(0.3$-$8.0 keV)} & \colhead{} \\
\colhead{(1)} & \colhead{(2)} & \colhead{(3)} &
\colhead{(4)} & \colhead{(5)}  }
\startdata
A1 & Field     &  TD          & 2.42          & BH-LMXB \\
A2 & GC        &  Canonical   & 0.79          & NS/BH-GC \\        
A4 & Field     &  HS          & 2.14          & NS/BH-LMXB \\
A5 & Field     &  QSS         & 9.42          & WD-WD binary \\
A6 & Field     &  SSS         & 1.12          & Classical Nova (CNe) \\
A7 & Field     &  Canonical   & 0.75          & NS/BH-LMXB \\
A8 & GC        &  Ionized HS  & 27.50         & GC-ULX in outflow \\
B1 & GC        &  TD          & 5.36          & BH-GC \\
B2 & Field   &  TD         & 1.03          & BH-LMXB \\
B3 & Field   &  TD         & 2.57          & BH-LMXB \\
B4 & Field   &  HS         & 1.22          & NS/BH-LMXB \\
B5 & Field   &  Canonical  & 0.63          & NS/BH-LMXB \\
B6 & Field   &  HS (+ disc contribution) & 23.22          & BH-LMXB \\
B7 & Field   &  HS         & 2.53          & NS/BH-LMXB \\
B8 & Field   &  QSS         & 1.11          & WD/NS-LMXB \\
C1 & Field   &  SSS+Gauss   & 3.40          & WD binary with \\
   &         &              &               & photospheric expansion \\
C2 & Field   &  Canonical   & 0.40          & NS/BH-LMXB \\
\enddata
\tablecomments{Column (3): TD indicates a source in a thermally dominant state, HS a source in a hard state and Canonical a source with insufficient counts to allow spectral modeling to be carried out. Details of the spectral modeling and interpretation of these sources are presented in Sections \ref{subsec:specanaly} \& \ref{sec:results} of this paper.}
\end{deluxetable}

\end{document}